\documentclass{elsart}

\usepackage{amsfonts}
\usepackage{epic,eepic}
\usepackage{graphicx}

\newcommand{\tr}{\mbox{\rm Tr}}

\newcommand{\bb}{{\bf{B}}}
\newcommand{\bm}{{\bf{M}}}
\newcommand{\bj}{{\bf {J}}}
\newcommand{\bt}{{\bf{T}}}
\newcommand{\bh}{{\bf{H}}}
\newcommand{\bw}{{\bf{W}}}
\newcommand{\drp}{\Delta_\Psi r^{2}(t)}
\renewcommand{\phi}{\varphi}
\renewcommand{\rho}{\varrho}
\renewcommand{\epsilon}{\varepsilon}
\renewcommand{\theta}{\vartheta}

\begin{document}

\begin{frontmatter}
\title {Crossover from Diffusive to Ballistic Transport in Periodic
  Quantum Maps}

\author{Daniel K. W\'{o}jcik\thanksref{email}\thanksref{homepage}} 
\thanks[email]{email: danek@cns.physics.gatech.edu}
\thanks[homepage]{www: http://www.cns.gatech.edu/\~{}danek/}
\address{Center for Nonlinear Science, School of Physics, Georgia
  Institute of Technology, 837 State Street, Atlanta, GA 30332-0430,
  USA} 
\address{Centrum Fizyki Teoretycznej Polskiej Akademii Nauk, Al.
  Lotnik\'{o}w 32/46, 02-668 Warszawa, Poland} 
\author{J. R. Dorfman}
\address{Institute for Physical Science and Technology, and
  Department of Physics, University of Maryland, College Park,
  Maryland 20742}


\date{\today}

\begin{abstract}
  We derive an expression for the mean square displacement of a
  particle whose motion is governed by a uniform, periodic, quantum
  multi-baker map. The expression is a function of both time, $t$, and
  Planck's constant, $\hbar$, and allows a study of both the long
  time, $t\rightarrow\infty$, and semi-classical, $\hbar\rightarrow
  0$, limits taken in either order. We evaluate the expression using
  random matrix theory as well as numerically, and observe good
  agreement between both sets of results.
  
  The long time limit shows that particle transport is generically
  ballistic, for any fixed value of Planck's constant.  However, for
  fixed times, the semi-classical limit leads to diffusion.  The mean
  square displacement for non-zero Planck's constant, and finite time,
  exhibits a crossover from diffusive to ballistic motion, with
  crossover time on the order of the inverse of Planck's constant.
  
  We argue, that these results are generic for a large class of 1D
  quantum random walks, similar to the quantum multi-baker, and that~a
  sufficient condition for diffusion in the semi-classical limit is
  classically chaotic dynamics in each cell.  Some connections between
  our work and the other literature on quantum random walks are
  discussed.  These walks are of some interest in the theory of
  quantum computation.
\end{abstract}

\end{frontmatter}

\section{Introduction}
It is well known that the quantum properties of classically chaotic
systems differ considerably from their classical
versions~\cite{berry79da,haake01da,stoeckmann99da}. The quantum
dynamics of a classically chaotic system with a finite number of
degrees of freedom is considerably more regular than its classical
counterpart, as indicated by the structures developed by Wigner
distributions in phase space for simple systems, and the fact that
almost all definitions of the quantum mechanical version of the
Kolmogorov-Sinai rate of entropy production give the value zero for
finite quantum systems that are classically chaotic with positive
Kolmogorov-Sinai entropy~\cite{alicki01da}. Furthermore, it is often
difficult to describe the behavior of quantum systems for both long
times and for small values of Planck's constant, {\it i.e.}  in the
semi-classical regime.  This difficulty is related to the fact that
the long time limit and the limit of small Planck constant do not
commute, in addition to analytical problems that often make explicit
solutions of the equations for quantum systems difficult.  Therefore
it is of some interest to study simple models where both the quantum
and classical properties are accessible to analytic and simple
numerical studies, and where the two limits can be studied in detail.
One example is the baker map.  The classical version of this map is
tractable analytically~\cite{hopf37da,dorfman99s,gaspard98sa}, and the
quantum version is less so~\cite{BalazsV89,Saraceno90}, but easily
studied using numerical methods.

It is even more challenging to study transport problems in detail. In
recent years some attention has been devoted to connecting macroscopic
transport properties with microscopic chaos, see {\em
  e.g.}~\cite{dorfman99s,gaspard98sa} and other contributions to this
volume.  We find it of interest to study the quantum signatures of
these relations and look for their appearance in the semi-classical
regime.  To gain intuition about the change in the character of
transport properties from quantum to classical regime a convenient
system to study is the uniform, periodic multi-baker map. Our interest
in this map is stimulated by the fact that the classical version is a
chaotic system with transport, in this case, diffusion, so it provides
a model where one can study transport in the context of chaotic
dynamics.

Both the classical~\cite{gaspard92s,tasaki95s} and
quantum~\cite{wojcik02da,wojcik02dd} versions of the multi-baker map
are based upon the baker map, which in its classical version is an
area preserving, expanding and contracting transformation of the
2-torus onto itself. The multi-baker map is obtained by considering a
two dimensional strip of unit height in the $y$-direction and a
segment of the real $x$-axis, either of infinite length in both
directions or of finite length with specified boundary
conditions~\cite{gaspard92s,tasaki95s}. The multi-baker transformation
is a combination of the baker map with a translation of points in each
unit interval to corresponding points in the nearest intervals to the
right or left according to a well defined prescription, to be given
below. The quantum versions of these maps are obtained by means of a
simple quantization where the $x$-axis is taken to correspond to
position space, and the $y$-direction is taken to correspond to
momentum space~\cite{wojcik02da,wojcik02dd}. The quantum mechanics is
obtained by requiring that there be an integral number of quantum
states in a unit interval, and that the time development of these
states correspond to the expanding and contracting properties of the
classical map.  Balazs and Voros~\cite{BalazsV89}, and
Saraceno~\cite{Saraceno90} were able to show that these requirements
can be satisfied by means of a unitary operation on the quantum
states, provided Planck's constant is taken to be the inverse of the
number of quantum states in the unit square. Our work has been devoted
to extending this quantum baker map to a quantum multi-baker map and
examining the transport properties of this quantum
map~\cite{wojcik02da,wojcik02dd}. Other quantizations of multibaker
maps as well as the Kapral-Elskens coupled baker model have been
studied in~\cite{LakshminarayanB94}

In this paper we focus on the calculation of the mean square
displacement (MSD) of a particle whose dynamics is described by a
quantum multi-baker map. This quantity is of importance for the
description of the average motion of a particle, since the time
dependence of this quantity can distinguish between sub-diffusive,
diffusive, and super-diffusive motion. A particular example of
super-diffusive motion is, of course, ballistic motion characteristic
of the motion of a free particle.

In the classical multi-baker, the MSD grows linearly in time, $\langle
(\Delta r)^2 \rangle=t$, characteristic of diffusion, for all times
greater than some microscopic time. The quantum version shows an
asymptotic ballistic, $t^2$, growth for what we will call uniform
maps, which are translationally invariant. This is essentially the
same as one finds in models of electron transport in periodic solids,
such as the Kr\"onig-Penney model~\cite{baym69f}. One can find
examples of non-translationally invariant multi-bakers where the
particle is localized with no asymptotic growth of the MSD with time.
Here we consider only translationally invariant models, and leave the
large class of non-translationally invariant models for future work.

The classical multi-baker map~\cite{gaspard92s,tasaki95s} is a simple
model for deterministic diffusion along a one-dimensional lattice. It
can easily be constructed to be isomorphic to a random walk with any
given probabilities for jumping to the right and to the left along the
lattice. The quantum multi-baker map~\cite{wojcik02da,wojcik02dd} is
also an example of a random walk, but in this case, the walk is
quantum mechanical and has very different properties from the
classical map due to the interference of probability amplitudes for
possible paths. The subject of quantum random walks has developed a
literature over the past few years.  Here we will also discuss the
connection of the quantum multi-baker with other models of quantum
walks. We mention now that most of the previous studies by other
authors have been devoted to the so-called Hadamard
walk~~\cite{ambainis01f}, which is a special case of the quantum
multi-baker for the largest allowed value of Planck's constant. Thus
the multi-baker map studied here represents a generalization of the
Hadamard walk to all allowed values of Planck's constant.

The plan of this paper is as follows. In Section~\ref{sec:qmb} we
present the basic equations defining the quantum multi-baker map and
compare it to the classical version. In Section~\ref{sec:msd} we
discuss the MSD for the quantum map, and in Section~\ref{sec:rmt} show
how random matrix theory may be used to evaluate the MSD for small
values of Planck's constant, leaving some technical details for a
longer paper~\cite{wojcik02dc}. In Section~\ref{sec:numerics} we
compare the results of random matrix theory with numerical studies of
the map, which shows that random matrix theory is quite effective in
reproducing the numerical results and in providing an expression
showing the crossover from diffusive to ballistic motion for fixed,
non-zero values of Planck's constant as the time becomes large. In
Section~\ref{sec:qrw} we discuss the connection between the
multi-baker map and other models of quantum random walks. We conclude
with a summary of results, and a discussion of the possible
implications and extensions of this work.

\section{The quantum multi-baker map}  
\label{sec:qmb}

We begin with the classical multi-baker
map~\cite{hopf37da,gaspard92s,tasaki95s}. It is a two-dimensional
lattice system where the phase space at each lattice site is a square
and the dynamics is a combination of transport of the phase space
densities to neighboring cells, which models the free flight, followed
by a local baker map evolution within a square, which models a
collision with a fixed scatterer. That is, the multi-baker map, $M$,
is a composition of two maps $M = B \circ T$: transport, $T$, of phase
points to neighboring cells, given by
\[
  T(n,x,y) =
  \left\{
    \begin{array}[c]{ll}
      (n+1, x, y), &\quad {\rm for} \; 0 \leq x < 1/2, \\
      (n-1, x, y), &\quad {\rm for} \; 1/2 \leq x < 1,
    \end{array}
  \right.
\]
and the baker map, $B$, which acts on the $x,y$ coordinates of each
cell, $n$, separately, according to
\[
  B(n,x,y) = 
  \left\{
    \begin{array}[c]{ll}
      (n, 2x, y/2),         &\quad {\rm for} \; 0 \leq x < 1/2, \\
      (n, 2x-1,  (1+ y)/2), &\quad {\rm for}\; 1/2 \leq x < 1.
    \end{array}
  \right.
\]
The combination of these two maps is the multi-baker map which is a
time-reversible, measure preserving, chaotic transformation, with
evolution law
\[
  M(n,x,y) =
  \left\{
    \begin{array}[c]{ll}
      (n+1, 2x, y/2), &\,{\rm for} \; 0 \leq x < 1/2, \\
      (n-1, 2x-1,  (1+ y)/2), &\, {\rm for}\; 1/2 \leq x < 1.
    \end{array}
  \right.
\]
As mentioned above, this classical map represents a simple
area-preserving model of simple random walk.  It can also be
considered a simplified Bernoulli map for the motion of light weakly
interacting particles in a gas of heavy scatterers on regular lattice
(periodic Lorentz gas).

The quantization of the multi-baker map is based upon the known
quantization of the baker map on the unit square, carried out by
Balazs and Voros, and by Saraceno~\cite{BalazsV89,Saraceno90}. We form
the quantum multi-baker by transporting some of the quantum states to
the next cell on the right and the others to the next cell on the
left, based upon the way the transformed points are moved to
neighboring cells in the classical multi-baker, and then model the
scattering by the quantum baker map $B$ acting in every single cell.

To quantize the baker map, we regard the horizontal direction of the
torus $[0,1)^2$ as the position axis, while vertical axis corresponds
to the momentum direction. To obtain the Hilbert space we take the
subspace of the wave functions on a line whose probability densities,
$|\Psi(x)|^2,|\tilde{\Psi}(p)|^2$ are periodic in both position and
momentum representations, respectively: $\Psi(x+1) = \exp(i2\pi
\phi_q) \Psi(x)$, $\tilde{\Psi}(p+1) = \exp(i2\pi \phi_p)
\tilde{\Psi}(p)$, where $\phi_q, \phi_p \in [0,1)$ are phases
parameterizing quantization.  The quantization of the baker map
requires the phase space volume to be an integer multiple of the
quantum of
action~\cite{BalazsV89,Saraceno90,Hannay80,DeBievreEG96,DeBievreE98,deBievre01f}.
Therefore the effective Planck constant is $h=1/N$, for a baker map on
a unit torus, where $N$, an integer, is the dimension of the Hilbert
space.  The space and momentum representations are connected by a
discrete Fourier transform $\langle p_k|q_l\rangle = [G_N(\phi_q,
\phi_p)]_{k,l} := {N}^{-1/2} \exp(-i 2\pi N p_k q_l)$. The discrete
positions and momenta are $q_l = (l+\phi_q)/N$, $p_k = (k+\phi_p)/N$,
respectively.  We take $N$ to be an even integer and label the quantum
states in coordinate space by $j$, where $j = 0,1,\ldots, N-1$. The
states $j = 0,\ldots N/2-1$ are called the ``left'' states, with
collective wave function, $\Psi_L$, while the remaining states, $j=
N/2,\ldots, N-1$ are called the ``right'' states, with wave function,
$\Psi_R$. One may think of $\Psi_{R,L}$ as $N/2$ component vectors.

Having constructed the Hilbert space, one looks for a family of
unitary propagators parameterized by $N=1/h$ which approach the
classical map in semi-classical limit.  The quantum baker
map~\cite{BalazsV89,Saraceno90,DeBievreEG96} is then given by a
combination of two operations. The first operation takes the $N/2$
``left'' states into momentum states labeled $k=0,1,\ldots, N/2 -1$,
called ``bottom'' momentum states, by means of the $N/2\times N/2$
Fourier transform matrix $G_{N/2}(\phi_q,\phi_p)$. The ``right''
coordinate states are transformed to ``top'' momentum states in the
same way.  Now one has transformed $N$ spatial states into $N$
momentum states in a way that mimics the classical baker's map. The
final step is to express the $N$ new states in coordinate
representation by means of the $N\times N$ matrix, $G_N^{-1}$, which
takes $N$ states in the momentum representation to their coordinate
representation. The full baker transformation, $\bb$, on the torus is
then given by the unitary transformation
\begin{equation}
\bb := G_N^{-1} \left[
  \begin{array}[c]{cc}
    G_{N/2} & 0 \\ 0 & G_{N/2}
  \end{array}
\right]
\label{BM}
\end{equation}
for even $N$.  Other examples and discussions of issues concerning the
quantization of area-preserving maps can be found for instance
in~\cite{BalazsV89,Saraceno90,Hannay80,DeBievreEG96,DeBievreE98,deBievre01f,baecker02f}.

Using this transformation as a template we can easily express the
quantum multi-baker map as a transformation of right and left quantum
states in unit squares labeled by $n\pm 1$ to right and left states in
the unit square labeled by $n$, $\Psi_{R,L}(n)$.  That is, in the
position representation the evolution of the wave function in the
quantum multi-baker is given by the equations\footnote{Note that the
  construction we use here differs slightly from the one we studied in
  our previous work~\cite{wojcik02da}. We change it to make the
  construction consistent with the general picture that we now have as
  well as to make model clearer. The properties of the two versions of
  random model should be the same, and the periodic model is
  unaffected.}
\begin{eqnarray}
  \label{eq:qmb}
  \left[
    \begin{array}[c]{c}
      \Psi_L(n,t+1) \\ \Psi_R(n,t+1)
    \end{array}
  \right]  
  & = & 
  G_N^{-1}(n) \cdot
  \left[
    \begin{array}[c]{cc}
      G_{N/2}(n) & 0\\ 
      0 & G_{N/2}(n)
    \end{array}
  \right]\cdot
  \left[
    \begin{array}[c]{c}
      \Psi_L(n-1,t) \\ \Psi_R(n+1,t)
    \end{array}
  \right] 
\end{eqnarray} 
where $G_N(n)$ is the discrete Fourier transform
\begin{equation}
  (G_N(n))_{kj} = \frac{1}{\sqrt{N}}e^{-2\pi i (k +
    \phi_p(n))(j+ \phi_q(n))/N}.  \label{eq:3}
\end{equation}
In principle, the Fourier transformation matrix $G_N(n)$ can depend
upon the cell index $n$. Here we will consider the case where this
matrix is independent of the cell index, so that the system of
equations is translationally invariant from cell to cell. We restrict
our attention here to finite systems of length $L$ with periodic
boundary conditions. This condition induces Bloch states and an
eventual ballistic motion of particles through the chain.  We
associate $L$-dimensional Hilbert space with the lattice, and since
the internal space is ${\mathbb C}^N$, the Hilbert space of the system
is the tensor product ${\mathbb C}^L \otimes {\mathbb C}^N$.  We will
work with the basis defined by $|n,\pm,i\rangle$, where $n$ is the
lattice site, $\pm$ denotes left/right half of the square, and
$i=0,\dots,N/2-1$ denotes the state in given half in the position
representation.  Since the states associated with the left half of the
cell go one step to the right we write $\Psi_L \equiv \Psi_+$,
accordingly $\Psi_R \equiv \Psi_-$. Then the general wave function for
the multi-baker chain can be written as
\begin{equation}
  |\Psi \rangle = \sum_n \sum_{\epsilon=\pm} |\Psi_{\epsilon}(n)\rangle,
\end{equation}
where $|\Psi_{\epsilon}(n)\rangle := P_{\epsilon}(n) |\Psi \rangle$,
$P_\epsilon(n):= \sum_i |n,\epsilon,i\rangle\langle n,\epsilon,i| $ is
orthogonal projection of the state vector onto the ``left'' or
``right'' subspace at site $n$. Thus the inner product takes form 
\begin{equation}
  \langle \Phi | \Psi \rangle := \sum_n \sum_{\epsilon=\pm}
  \langle \Phi_{\epsilon}(n) | \Psi_{\epsilon}(n)\rangle.
\end{equation}
We may think of $|\Psi_{\epsilon}(n)\rangle$ and $\langle
\Psi_{\epsilon}(n)|$ as corresponding to $N/2$-component column and
row vectors, respectively, while $|\Psi\rangle$ corresponds to an
$NL$-com\-po\-nent vector.
Since we work mostly in position basis, 
we use $N/2$ dimensional vectors $\Psi_{\epsilon}(n)$, whose $i$-th
components are $\Psi_{\pm}(n,i)= \langle n,\pm,i | \Psi\rangle$, where
the index $n$ denotes the lattice site of the cell under
consideration, as above, the notation $\pm$ denotes the particular
half of the unit square under consideration, the left half indicated
by $+$ moves to the cell to the right, and the right half, denoted by
$-$, moves to the cell to the left. The index $i$ denotes a particular
quantum state in the right or left half of the unit cell, specified by
$\pm$ and $i=0,1,\ldots, N/2 -1.$ The action of the quantum
multi-baker map on the total wave function will be written as the
action of an operator $\bm$ defined by Eq.  (\ref{eq:qmb}), as
\begin{equation}
|\Psi'\rangle = \bm|\Psi\rangle.
\label{jrd1}
\end{equation}   
The operator $\bm$ is the Floquet operator for the multi-baker map and
will be used to determine the time dependence of various observables
for this system, as
\begin{equation}
\Omega(t) = {\bm}^{\dagger\,t}\Omega {\bm}^{t},
\label{jrd2}
\end{equation}
where $\Omega$ is any observable of the system.

The complete specification of the model requires values for the phases
$\phi_q,\phi_p$. There is a considerable amount of freedom for
choosing these phases. As mentioned above, one may choose them to vary
from cell to cell.  A random variation of phases from cell to cell
produces a {\it disordered} multi-baker map, while requiring that the
phases have constant values throughout the lattice produces a {\it
  regular, or uniform} multi-baker map which is translationally
invariant from one cell to the next. Many other choices are possible
and can be of interest. Here we consider only the translationally
invariant case. In numerical calculations we use the values for the
phases chosen by Balazs and Voros~\cite{BalazsV89} with $\phi_q =
\phi_p =0$; those used by Saraceno~\cite{Saraceno90} $\phi_q = \phi_p
= 1/2$, which lead to survival of additional classical symmetry in the
model, as well as more generic values. We will see below that the
Saraceno phases lead to a non-generic behavior for the MSD.

Let us consider now the structure of eigenstates of the unitary
operator $\bm$. Since we consider periodic boundary conditions every
eigenstate corresponding to the eigenvalue $\lambda=e^{i\kappa}$ has a
Bloch form
\begin{eqnarray}
  \Psi(n,\pm) & = & \exp(i \kappa n) \tilde{\Psi}_{\pm} / \sqrt{L}
  \label{eq:ansatz}
\end{eqnarray}
where
$  \left [ 
    \begin{array}[c]{c}
      \tilde{\Psi}_+ \\
      \tilde{\Psi}_-
    \end{array}
  \right ]$
  is the normalized eigenstate of a modified quantum baker operator
\begin{equation}
  G_N^{-1}
  \left[ 
    \begin{array}[c]{cc}
      G_{N/2} e^{-i\kappa}& 0 \\
      0 & G_{N/2} e^{i\kappa}
    \end{array}
  \right].
\label{9}
\end{equation}
Periodic boundary conditions imply $e^{i \kappa L} = 1$, thus we have
$\kappa_k = 2\pi k/L, k=0, 1, \dots, L-1$.  For every $\kappa_k$ we
have $N$ eigenstates. We will enumerate the eigenstates of $M$ by
$k,n$, with $n=0, \dots, N-1$. Thus $\Psi_n^k$ is the eigenstate given
by the ansatz~(\ref{eq:ansatz}) with $\kappa_k = 2\pi k/L$,
corresponding to the eigenvalue $e^{i \phi_n^k}$, where phases
$\phi_n^k$ are counted for a given $k$ from 0 (including) to $2 \pi$,
i.e. $0 \leq \phi_n^k \leq \phi_{n+1}^k < 2\pi$.

\section{The mean square displacement (MSD) in the uniform multi-baker map}
\label{sec:msd}

In order to formulate our calculation of the MSD for the uniform
multi-baker map in the most convenient way, we introduce two
operators, $r,v$, which represent a coarse position operator and a
coarse velocity operator, respectively. These operators are coarse in
the sense that $r$ simply gives the lattice site associated with a
particular quantum state, and the coarse grained velocity $v$ is given
by $v= \bm^\dagger r \bm -r$. We argue elsewhere~\cite{wojcik02dc},
that for a translationally invariant system $v$ has a very simple form
with values $\pm 1$, given by the change in cell index for each
quantum state. Thus
\begin{eqnarray}
r|n,\pm,i\rangle & = & n|n,\pm,i \rangle \nonumber \\
v|n,\pm,i\rangle & = & \pm|n,\pm,i\rangle,
\label{jrd3}
\end{eqnarray}
so that explicit expressions for the operators $r,v$ are
\begin{eqnarray}
  r & = & \sum_{n,\epsilon,i} n |n,\epsilon,i \rangle \langle
    n,i,\epsilon|,\nonumber \\ 
  v & = & \sum_{n,\epsilon,i} \epsilon |n,\epsilon,i \rangle 
  \langle n,\epsilon, i|.
  \label{jrd4}
\end{eqnarray}  
Suppose we prepare the system in a pure state $|\Psi\rangle$. Then the
mean square displacement of the particle starting in this state is
given by
\begin{eqnarray}
  \langle (\Delta r)^2(t)\rangle_\Psi 
  = \langle (\bm^{\dagger t} r \bm^t -r)^2 \rangle_\Psi 
  = \langle (\sum_{\tau=0}^{t-1} v_\tau)^2 \rangle_\Psi 
  = \sum_{\tau_1,\tau_2 =0}^{t-1} \langle v_{\tau_1} v_{\tau_2}\rangle_\Psi,
\end{eqnarray}
where $\langle A \rangle_\Psi := \langle \Psi | A | \Psi \rangle =
\tr\ (| \Psi \rangle\langle \Psi | A) $. Depending on the original
state we have a distribution of possible results. To characterize it
we can calculate its average over all the possible initial states, and
the root mean square deviation from the average, which quantifies the
spread of the results, or the quality of the prediction based on the
average. In this Section we find the expressions for both the
equilibrium MSD as well as for the equilibrium fluctuations of this
function. In the next section we approximate the average results using
random matrix theory, and then we compare them with numerical
evaluation of the exact formulas.

Since the average over all the possible pure states gives the most
incoherent mixture,
we obtain a simple expression for the equilibrium MSD as
\begin{eqnarray}
  \langle (\Delta r)^2(t)\rangle
  = \sum_{\tau_1,\tau_2 =0}^{t-1} \langle v_{\tau_1} v_{\tau_2}\rangle,
\end{eqnarray}
where
\begin{equation}
  \langle A \rangle := \tr\ (\rho_{\rm eq} A) = \frac{1}{LN} \tr\ (A),
\label{tra}
\end{equation}
$L$ is the length of the chain (assume periodic boundary conditions),
and $N$ is the dimension of the Hilbert space for a single cell.
Clearly, $\rho_{\rm eq} = 1_{NL}/(LN) $ is the equilibrium state for
all the QMB, that is, it is time invariant
\begin{equation}
  \bm \rho_{\rm eq} \bm^\dagger = \rho_{\rm eq}
\end{equation}
and it maximizes the entropy $S = -\tr\ (\rho \ln \rho)$.  Time
invariance implies that the velocity autocorrelation function
\begin{equation}
  C_{\tau_1,\tau_2} := \langle v_{\tau_1} v_{\tau_2} \rangle = \langle
    \bm^{\dagger\tau_1} v \bm^{\tau_1} \bm^{\dagger \tau_2} v
    \bm^{\tau_2} \rangle     
\end{equation}
satisfies
$ C_{\tau_1,\tau_2} = C_{\tau_1-\tau_2,0} = C_{0,\tau_1-\tau_2} \equiv
  C_{\tau_1-\tau_2}.$
Thus we can write
\begin{eqnarray}
  \langle (\Delta r)^2(t)\rangle & = & \sum_{\tau=0}^{t-1} 
  C_0 
  + 2 \sum_{\tau_1>\tau_2=0}^{t-1} 
  C_{\tau_1-\tau_2} 
  =  
  C_0 t+ 2 \sum_{\tau=1}^{t-1}
  (t-\tau) C_{\tau} .
  \label{gk}
\end{eqnarray}
This leads to a Green-Kubo like formula for the diffusion coefficient
\begin{eqnarray*}
  D & = & \lim_{t \rightarrow \infty} \frac{1}{2t} \langle (\Delta
  r)^2(t)\rangle = \frac12 C_0 
  + \sum_{\tau=1}^\infty C_\tau   
  = \sum_{\tau=0}^\infty  
  C_\tau -\frac12 C_0, 
\end{eqnarray*}
paralleling the classical case. However, here the diffusion
coefficient turns out to be infinite.
 
Thus the MSD is written as the sum of time correlations of the
velocity, just as in the classical case, but the difference in
mechanics will lead to important differences in the time development
of the MSD.

The translational invariance of the system, together with the uniform
equilibrium density, imply that instead of summing over all the $N$
quantum states in the $L$ cells, we can sum only over the $N$ quantum
states in one cell, see Ref. \cite{wojcik02dc}. Thus we can write
\begin{eqnarray}
  C_\tau  & = & \frac{1}{LN}\tr\
  [\bm^{\dagger \tau} v \bm^\tau v ] \nonumber \\
  & = & \frac{1}{N} \sum_{i,\epsilon=\pm}
  \langle n= 0, \epsilon, i| \bm^{\dagger \tau} v \bm^\tau v  |n= 0,
  \epsilon, i\rangle \nonumber \\
  & = & \frac{1}{N} \sum_{i,\epsilon=\pm} \epsilon 
  \langle n=0, \epsilon, i| \bm^{\dagger \tau} v \bm^\tau   |n= 0,
  \epsilon, i\rangle. 
\label{jrd5}
\end{eqnarray}
Thus we have to sum, with appropriate signs, over all the possible
cyclic paths starting in a fixed cell. The next simplification
provided by the translational invariance of the model is that we can
replace the unitary operator $\bm$ for the full multi-baker in Eq.
(\ref{jrd5}) by the operator $\bb$ acting within a unit cell. This
replacement follows from the observation that the velocity operator is
diagonal in the representation based on the states in the cells, with
a block structure that is identical from one cell to the
next~\footnote{This would not be the case for the disordered
  multi-baker maps where the quantization phases vary from one cell to
  the next.}.  Therefore, the velocity operator at time step $t$ takes
on the same value in all of quantum states that are periodic images of
any state in a fixed cell.  Thus we can express the velocity
correlation function, $C_\tau$, as
\begin{equation}
  C_\tau = \frac1N \tr\ [\bb^{\dagger \tau} \bj \bb^\tau \bj]
\label{jrd6}
\end{equation}  
where the matrix $\bj$ is the velocity operator restricted to a single
cell, in position representation given by
\begin{equation}
  \bj = \left [ 
    \begin{array}[c]{cc}
      1_{N/2} & 0 \\
      0 & -1_{N/2}
    \end{array}
  \right ].
\label{jrd7}
\end{equation}
Here $1_{N/2}$ is the $N/2\times N/2$ unit operator.  

To proceed further we must now use specific properties of the quantum
baker map $\bb$, and in particular, its spectral properties. Since
$\bb$ is a unitary operator, its eigenvalues lie on the unit circle
and define a set of $N$ phases, $\phi_j$ or quasi-energies. We will
denote the corresponding eigenstates by the Dirac kets $|j\rangle$.
Thus the spectral problem for the operator $\bb$ is
\begin{equation}
  \bb | j\rangle = e^{i \phi_{j}} | j\rangle .
\label{jrd8}
\end{equation} 
Then we can express the velocity correlation function $C_\tau$ as
\begin{eqnarray}
  C_\tau & = & \frac1N \sum_{j,k} e^{-i\phi_k \tau} \langle k | \bj | j \rangle
  e^{i \phi_j \tau} \langle j | \bj | k \rangle\nonumber \\
  & = & \frac1N \sum_{j,k} |\bj_{jk}|^2  e^{i (\phi_j-\phi_k) \tau}\\\nonumber
  & = & \frac1N [\sum_{j} |\bj_{jj}|^2 + 2\sum_{j>k}|\bj_{jk}|^2  \cos
  (\phi_j-\phi_k) \tau]. 
\label{jrd9}
\end{eqnarray}
The matrix elements of $\bj$ satisfy
\begin{equation}
  \sum_{j,k} |\bj_{jk}|^2 = \tr\ \bj^2 = N = \sum_j |\bj_{jj}|^2 +
  \sum_{j>k} 2 |\bj_{jk}|^2.
  \label{eq:23}
\end{equation}
Therefore
\begin{eqnarray}
  C_\tau & = & \frac1N\left[N + \sum_{j>k}|\bj_{jk}|^2 (e^{i
    (\phi_j-\phi_k) \tau} + e^{-i (\phi_j-\phi_k) \tau} -2)\right]  \nonumber \\
  & = & 1 - \frac4N\sum_{j>k}|\bj_{jk}|^2 \sin^2
  \frac{(\phi_j-\phi_k)\tau}{2}
\end{eqnarray}
We can now substitute these results into Eq. (\ref{gk}) for the mean square
displacement to obtain
\begin{eqnarray}
  \langle (\Delta r)^2(t) \rangle & = & t + \sum_{\tau=1}^{t-1}
  (t-\tau) C_\tau   \nonumber \\
  & = & t + t(t-1) \frac1N\sum_{j}|\bj_{jj}|^2  +
  \frac4N\sum_{j>k}|\bj_{jk}|^2 \sum_{\tau=1}^{t-1} \tau 
  \Re e^{i (\phi_j-\phi_k)(t-\tau)}
\label{jrd10}
\end{eqnarray}
The last expression is particularly useful for averaging and will be
used in the next section. 
 
The sums over the intermediate times $\tau$ in Eq. (\ref{jrd10}) can
easily be carried out. We assume now that there is no degeneracy in
the spectrum of quasi-energies and, after some algebra, we obtain an
expression for the MSD as
\begin{eqnarray}
  \langle (\Delta r)^2(t) \rangle & = &
  t^2\frac1N\sum_{j}|\bj_{jj}|^2  + \frac2N\sum_{j>k}|\bj_{jk}|^2
  \frac{\sin^2 \frac{(\phi_j-\phi_k) t}{2}}{\sin^2
    \frac{\phi_j-\phi_k}{2}} \label{eq:26} \\
  & = & \frac1N\sum_{j,k}|\bj_{jk}|^2 \frac{\sin^2
    \frac{(\phi_j-\phi_k) t}{2}}{\sin^2  \frac{\phi_j-\phi_k}{2}}.
  \label{jrd11}
\end{eqnarray}
Whenever two eigenphases $\phi_j,\phi_k$ are equal, the contribution
to the sum $\frac{\sin^2 \frac{(\phi_j-\phi_k) t}{2}}{\sin^2
  \frac{\phi_j-\phi_k}{2}}$ must be replaced by $t^2$.

We see that there is typically a ballistic contribution coming from
diagonal and possibly some degenerate terms. The other contributions
are oscillatory and usually negligible in the long time limit. It can
happen, however, that the ballistic contribution disappears altogether
and we have only oscillations which means that the particle is
localized. An example is provided in Section~\ref{sec:numerics} and an
explanation in more general context is given in the final Section. It
is interesting, that the obtained result is completely independent of
the length of the system. This is not true for the fluctuations of the
MSD.

It is not easy to evaluate this expression, Eq. (\ref{jrd11}),
analytically. Therefore in section~\ref{sec:rmt} we present an
approximation based on the random matrix theory, and in
Section~\ref{sec:numerics} we evaluate it numerically for some
particular choices of parameters, and compare it with the RMT
averages.

\subsection{Fluctuations in the MSD}

We have computed the MSD as an equilibrium average where all
eigenstates have the same weight. One might ask about the dependence
of the average square displacement for an individual quantum state. In
order to say something about the fluctuations of the average square
displacement from one state to the next, we now consider, $\drp$ the
average square displacement for the quantum state $|\Psi\rangle$,
where $\drp = \langle \Psi|\Delta r^{2}(t)|\Psi\rangle$.  We will try
to characterize the fluctuations of the square displacement among
various quantum states by calculating, in so far as it is possible,
the mean square fluctuation of $\drp$, which we denote by
$\Delta^{2}(t)$, where
\begin{equation}
  \Delta^{2}(t) =
  \left\langle \left[ \drp-\langle \Delta
    r^2(t) \rangle \right]^{2} \right\rangle_\Psi,
  \label{drp1}
\end{equation}
The average $\left\langle A(\Psi) \right\rangle_\Psi$ is defined as
follows: if $|\Psi_\alpha\rangle$ is an arbitrary orthonormal basis
and $b_\alpha := \langle \Psi_\alpha | \Psi\rangle$ are the complex
coefficients of expansion of $|\Psi\rangle$ in this basis then
\begin{equation}
  \left\langle A(\Psi) \right\rangle_\Psi := \frac{\int\!\! d^2 b_1
  \dots d^2 b_{LN} \, \delta(1 - \sum_{\alpha=1}^{NL} |b_\alpha|^2)
  A(\Psi)}{\int\!\! d^2 b_1 \dots d^2 b_{LN} \, \delta(1 -
  \sum_{\alpha=1}^{NL} |b_\alpha|^2)} = \frac{\int d\Omega_{2LN}
  A(\Psi)}{\int d\Omega_{2LN}} .   \label{eq:29}
\end{equation}
Thus it is an average over $2LN$-dimensional sphere.

Before we calculate the right hand side of Eq. (\ref{drp1}), we
note that the equilibrium mean square displacement can be written in a
very similar way as an average over the space of all the states 
\begin{equation}
\langle(\Delta r(t))^{2}(t) \rangle =
\langle\drp\rangle_\Psi.
\label{p1}
\end{equation}
Indeed, using Eq.~(\ref{eq:29}) we obtain 
\begin{equation}
  \langle(\Delta r(t))^{2}(t) \rangle =
  \sum_{\alpha,\beta} \langle b_{\alpha}^\ast b_{\beta} \rangle
  \langle \psi_{\alpha}|\Delta r^2(t)|\psi_{\beta}\rangle
  \label{p2}
\end{equation}
Only contributions from $\alpha=\beta$ survive due to phase averaging,
and the diagonal terms carry the same contribution, 
therefore
\(
\langle  b_\alpha^\ast b_{\beta} \rangle_\Psi 
=\delta_{\alpha,\beta}/(NL).
\)
If we insert this relation into the right hand side of Eq. (\ref{p2}),
we recover Eq. (\ref{tra}) with the quantity $A=\Delta r^2(t)$.

Returning to Eq. (\ref{drp1}), we see that the right hand side can be
written as
\begin{eqnarray}
  \Delta^{2}(t) & = & \sum_{n,m,k,l=0}^{t-1}
  \sum_{\alpha,\beta,\gamma,\delta=1}^{LN} 
  \langle b_\alpha^\ast b_\beta b_\gamma^\ast b_\delta \rangle_\Psi \langle 
  \Psi_\alpha | 
  v_n v_m | \Psi_\beta \rangle \langle \Psi_\gamma | v_k v_l |
  \Psi_\delta \rangle \nonumber \\
& & -\left[ \langle \Delta r^2(t) \rangle\right]^2.
\end{eqnarray}
Observe that if at least one phase, or equivalently one subscript, is
unpaired, the average $ \langle b_\alpha^\ast b_\beta b_\gamma^\ast
b_\delta \rangle$ will be zero. Therefore the average of the four
coefficients is non-zero only if the indices are paired. This can be
arranged in three ways, so that
\begin{equation}
  \langle b_\alpha^\ast b_\beta b_\gamma^\ast b_\delta \rangle = c_1
  \delta_{\alpha\beta} \delta_{\gamma\delta} (1 -
  \delta_{\beta\gamma}) + c_1 \delta_{\alpha\delta}
  \delta_{\gamma\beta} (1 - \delta_{\alpha\beta}) + c_2
  \delta_{\alpha\beta} \delta_{\gamma\delta} \delta_{\alpha\gamma}, 
\end{equation}
where
\begin{eqnarray}
  c_1 & = & \langle |b_1|^2 |b_2|^2 \rangle \\
  c_2 & = & \langle |b_1|^4 \rangle 
\end{eqnarray}
Let
$  A= \sum_{m,n=0}^{t-1} v_m v_n$.
Then 
\begin{eqnarray}
  \Delta^{2}(t) & = & c_1 \sum_{\alpha\neq \beta} \langle
  \Psi_\alpha | A |\Psi_\alpha\rangle  \langle
  \Psi_\beta | A |\Psi_\beta\rangle 
  + c_1 \sum_{\alpha\neq \beta} \langle
  \Psi_\alpha | A |\Psi_\beta\rangle  \langle
  \Psi_\beta | A |\Psi_\alpha\rangle \nonumber\\
  & & + \sum_\alpha c_2 \langle
  \Psi_\alpha | A |\Psi_\alpha\rangle  \langle
  \Psi_\alpha | A |\Psi_\alpha\rangle - \left[\frac{1}{NL}\tr
  A\right]^2 \nonumber \\ 
  & = & c_1 (\tr A)^2 + c_1 \tr A^2 + \sum_\alpha (c_2 - 2c_1)
  |\langle \Psi_\alpha | A |\Psi_\alpha\rangle|^2 -
  \left[\frac{1}{NL}\tr A\right]^2. 
\end{eqnarray}
Since the term $\sum_\alpha |\langle \Psi_\alpha | A
|\Psi_\alpha\rangle|^2 $ is basis dependent we must have $c_2= 2c_1$.
Also,
\begin{equation}
\sum_{\alpha,\beta}|b_{\alpha}|^2|b_{\beta}|^2
=\langle\sum_{\alpha,\beta}|b_{\alpha}|^2|b_{\beta}|^2\rangle =1.
\end{equation}  
Since there are $LN$ quantum states, this reduces to
\begin{equation}
  c_2 LN 
  + c_1 LN(LN-1) 
  = 1,
\end{equation}
therefore 
\begin{equation}
  c_1 = \frac{1}{LN (LN+1)} \qquad c_2 = \frac{2}{LN (LN+1)}.
\end{equation}

\begin{figure}[htbp]
  \centering
  \includegraphics[scale=.5]{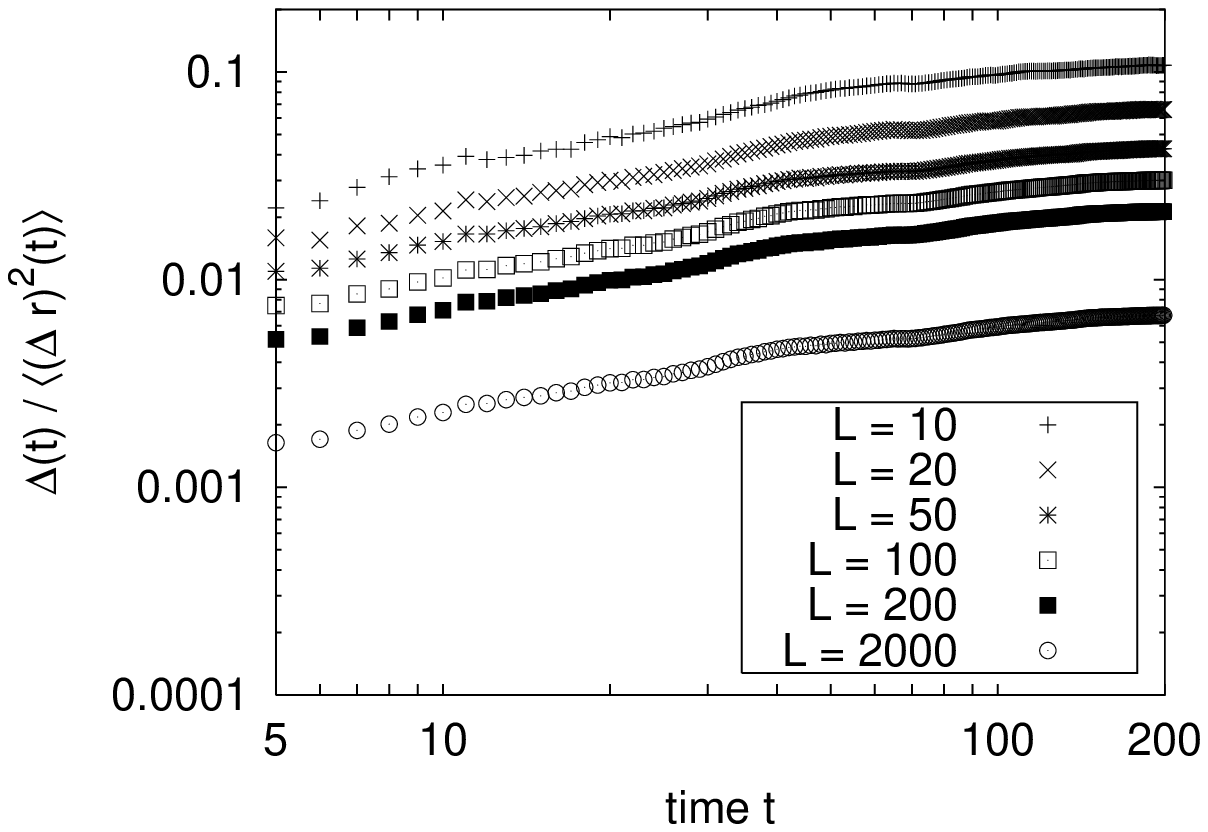} 
  \includegraphics[scale=.5]{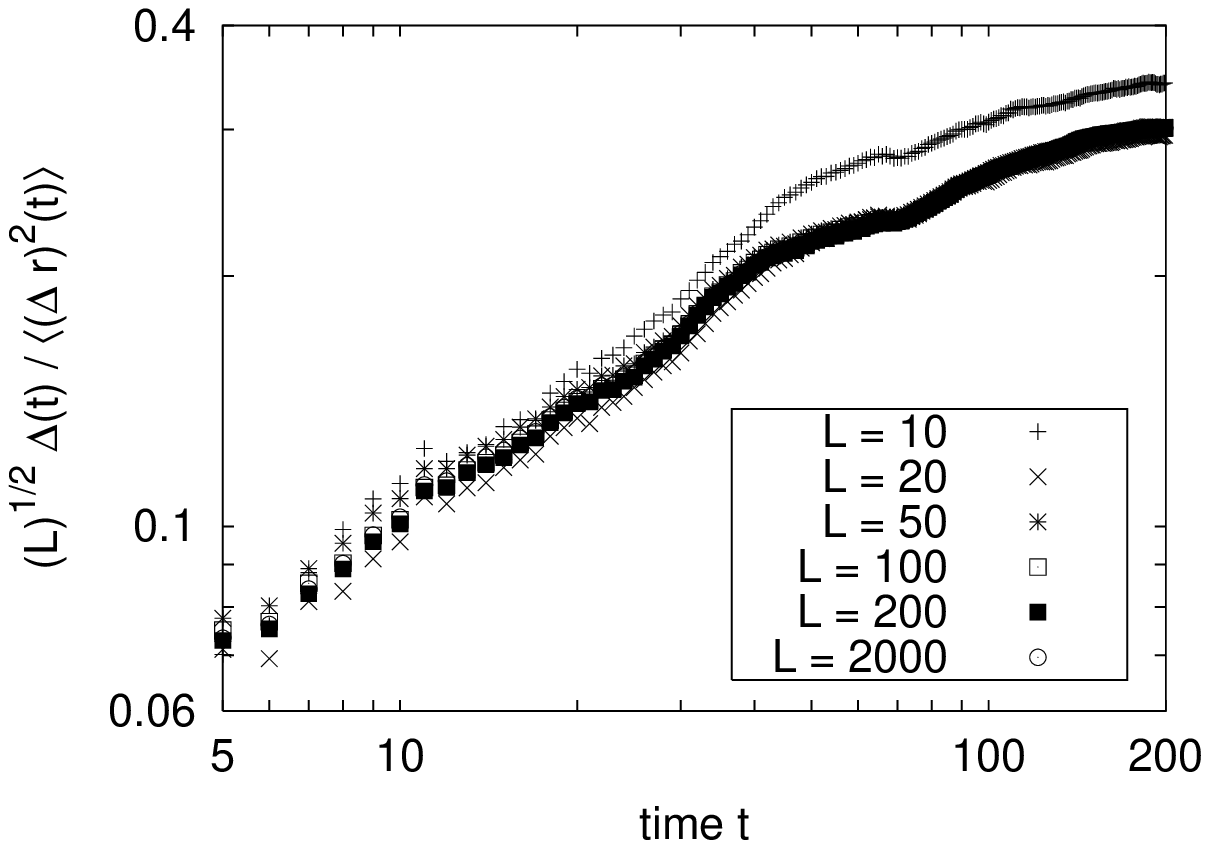}
  \caption{Time and length dependence of the relative fluctuations
    $\frac{\sqrt{\Delta^2 (t)}}{\langle \Delta r^2(t) \rangle}$. The
    left picture shows log-log plot of the unscaled data, the right
    picture shows each curve multiplied by square root of $L$. Plots
    made for $N=50, \phi_q = \phi_p =0$. }
  \label{fig:fluct}
\end{figure}

To proceed further we need to use the results obtained earlier
following Eq. (\ref{9}) for the eigenstates and eigenphases of the
multi-baker operator. The eigenstates have the form of periodic Bloch
states and the eigenphases appear in $L$ sets of $N$ eigenphases.
After some lengthy, but straightforward, calculations one finds the
final result for the fluctuations of the average square deviations
as~\cite{wojcik02dc}
\begin{eqnarray}
  \Delta^{2}(t) & = & c_1
  [(\tr\ A)^2 + \tr\ A^2]  - \frac{1}{L^2 N^2} (\tr\ A)^2 \nonumber \\
  & = & c_1 \left[ \tr\ A^2 - \frac{1}{LN} (\tr\ A)^2 \right] \nonumber \\
  & = & \frac{1}{L+1/N} \frac{1}{N^2} \left[
    \left\langle \sum_{p,s=0}^{N-1} 
  \left|
    \sum_{r=0}^{N-1} J_{pr}^a J_{rs}^a  
    \frac{\sin \frac{(\phi_p^a-\phi_r^a)t }{2}}{\sin
      \frac{\phi_p^a-\phi_r^a }{2}} 
    \frac{\sin \frac{(\phi_r^a-\phi_s^a)t}{2}}{\sin
      \frac{\phi_r^a-\phi_s^a }{2}} 
  \right|^2 \right\rangle_a \right. \nonumber \\
  & & \left.-\frac1N \left(\sum_{m,n=0}^{N-1} |J_{nm}|^2 \frac{\sin^2
  \frac{(\phi_n-\phi_m)t }{2}}{\sin^2\frac{\phi_n-\phi_m }{2}}
  \right)^2\right] 
\label{drp4}
\end{eqnarray}
The main source for the system size, $L$, dependence of this
fluctuation result is the factor of $L^{-1}$ in front of the whole
expression. This provides the usual $L^{-1/2}$ decay of the root mean
square fluctuation which suggests that in the large systems the
equilibrium average is typical for almost every initial, pure state.
The eigenphases $\phi^a_j$ also depend on the system size, so that in
the absence of a further evaluation of the sums, one cannot precisely
determine the size dependence of $\Delta^{2}(t)$, however, it is
reasonable to expect that the contributions for different Bloch
vectors are of the same order.  Numerical results suggest that this
expectation is correct for the quantum multi-bakers, and the final
result for the right hand side of Eq.~(\ref{drp4}) decays inversely
with $L$. This is shown in Figure~\ref{fig:fluct}. The left picture
shows the time and size dependence of the relative fluctuations
$\frac{\sqrt{\Delta^2 (t)}}{\langle \Delta r^2(t) \rangle}$ on a
log-log plot. The right picture shows the same data with each curve
multiplied by square root of $L$. The scaling is remarkable. Plots
were obtained through numerical evaluation of exact
formula~(\ref{drp4}) for $N=50, \phi_q = \phi_p =0$.
Using the methods of random matrix theory discussed in the next
section, in the limit of large $N$, one can replace the average over
the Bloch states, labeled by the subscript $a$, by averages over
random matrix ensembles. We have not yet done this and leave it for
future work.

\section{Random matrix theory for the mean square displacement} 
\label{sec:rmt}

To proceed further we need to know the quasi-energies and eigenvectors
of the operator $\bb$. For large $N$ it is not possible to determine
these quantities by analytical means. Instead one uses numerical
methods to diagonalize $\bb$. The results obtained this way will be
discussed in the next section. Here we will show how random matrix
theory~\cite{haake01da,Mehta91da} may be used to evaluate the
approximate value of the MSD. In a sense, the use of random matrix
theory must be considered to be a kind of uncontrolled approximation
method since there are no theorems indicating that the distributions
of quasi-energies are described by any of the three random matrix
ensembles, and there are no similar results for the eigenvectors.
However, since the classical baker map is an example of a strongly
chaotic system, we can expect that one of the two ensembles, the
circular orthogonal ensemble, COE, or the circular unitary ensemble,
CUE, might provide a very good approximation when it is used to
evaluate the right hand sides of Eq.~(\ref{jrd11}), see
ref.~\cite{Saraceno90,OconnorT91}. The results obtained this way will,
of course, be averages over the ensemble of random matrices, and not
characteristic of any one matrix in the ensemble, and perhaps not of
the baker map itself. We will see shortly that the numerical results
for the baker map are in very good agreement with the results of
random matrix theory, so the connection between the use of random
matrix methods for classically chaotic systems is supported by this
work.

We begin the application of random matrix theory by determining the
average values of the moduli of the matrix elements
$\bj_{jk}=\langle~j~|~\bj~|~k~\rangle$ where $|j\rangle,|k\rangle$
indicate eigenvectors of the unitary baker operator $\bb$.  We begin
with the diagonal elements, which we first express in the position
representation of the $N$ quantum states defined in the baker cell.
That is, we write
\begin{equation}
  |J_{jj}|^2 = | \langle j | J | j \rangle |^2 =
   \left(\sum_{\alpha=0}^{N/2-1} |\langle \alpha | j\rangle |^2 -
   \sum_{\alpha=N/2}^{N-1} |\langle \alpha | j\rangle |^2 \right)^2,
\label{jrd12}
\end{equation}
where the kets $|\alpha\rangle$ are basis vectors in position space. 
Since the eigenkets are normalized, that is,
$  \sum_{\alpha=0}^{N-1}  |\langle \alpha | j\rangle |^2 = 1$,
we may set $s_j = \sum_{\alpha=0}^{N/2-1} |\langle \alpha |
j\rangle |^2$ to obtain
\begin{equation}
  |J_{jj}|^2 = (2s_j-1)^2 = 4 s_j^2 - 4s_j + 1.
\end{equation}
To get the random matrix theory average value, we observe that the
joint probability of single eigenstate components in the CUE and COE
ensembles are given by~\cite{haake01da,Kus88f}
\begin{equation}
  P(\{ \Psi \}) = c \delta(1 - |\Psi|^2) = c
  \delta(1-\sum_{\alpha=0}^{2M-1} |\Psi_\alpha|^2).
\end{equation}
In the COE ensemble the eigenstates can be chosen real which implies
that $P(\{ \Psi \})$ is a uniform distribution on $N$-dimensional unit
sphere, and $M=N/2$. In CUE ensemble the eigenstates are complex
implying that $P(\{ \Psi \})$ is uniform on a $2N$-dimensional unit
sphere $(|\Psi_j|^2 = (\Re \Psi_j)^2 + (\Im \Psi_j)^2)$, and $M=N$.
These averages can be calculated in straightforward ways and one
obtains $\langle s \rangle = 1/2$ for both ensembles. However the
averages $\langle s^2 \rangle$ differ in the two ensembles. We obtain
\begin{equation}
  \langle s^2 \rangle =
  \frac{\Gamma(\frac{M}{2}+2)}{\Gamma(\frac{M}{2})}
  \frac{\Gamma(M)}{\Gamma(M+2)} = \frac{M+2}{4(M+1)}, 
\end{equation}
where $M=N/k$ with $k=2$ for the COE and $k=1$ for the CUE. This leads
to an evaluation of the averages needed for the diagonal terms, with
the result that
\begin{equation}
  \langle |J_{jj}|^2 \rangle = \frac{k}{N+k}.
  \label{jrd14}
\end{equation}
To get the average of the off-diagonal terms, we use~(\ref{eq:23}) 
which, after averaging, leads to
\begin{equation}
  \langle |J_{j\neq k}|^2 \rangle = \frac{N}{(N-1)(N+k)},
\label{jrd15}
\end{equation}
with $k$ given above.

\subsection{The MSD in the CUE ensemble}

Next we calculate the average of $e^{i (\phi_1 - \phi_2)\tau}$ in CUE
ensemble. To do this we need an expression for the pair correlation
function, $R(\phi_j,\phi_k) $ of two angles in this ensemble. This is
calculated in some detail by Mehta~\cite{Mehta91da}, and we use the
expression given there.
\begin{eqnarray}
  \langle e^{i (\phi_1 - \phi_2) \tau}\rangle & = & \int_0^{2\pi} d\phi_1
  \int_0^{2\pi} d\phi_2 e^{i (\phi_1 - \phi_2) \tau}
  \frac{R(\phi_j,\phi_k)}{N(N-1)} \nonumber \\
  & = & \int_0^{2\pi} d\phi_1
  \int_0^{2\pi} d\phi_2 e^{i (\phi_1 - \phi_2) \tau}
  \frac{N}{4 \pi^2 (N-1)}\left[1 -
    \frac{\sin^{2}\frac{N(\phi_j-\phi_k)}{2}}{N^2
      \sin^2\frac{(\phi_j-\phi_k)}{2}}\right].
\end{eqnarray}
At $\tau=0$ the first term is $N/(N-1)$, and it vanishes for $\tau
>0$. We calculate the second term by converting the angular integral
to one in the complex plane. By changing the variables to $u=\phi_1 -
\phi_2, v = \phi_2$, and then setting $z=e^{iu}, dz = i z\, du$, we
obtain, for $\tau>0$,
\begin{eqnarray}
  \langle e^{i (\phi_1 - \phi_2) \tau}\rangle 
  &=& \frac{-1}{2 \pi N(N-1)} \int_0^{2\pi} du \, e^{i u  \tau}\,
  \frac{\sin^{2}\frac{N u}{2}}{
    \sin^2\frac{u}{2}} \\
  & = & \frac{-1}{2 \pi i N(N-1)} \oint dz\, z^{\tau-N} g(z),
\end{eqnarray}
where 
\begin{equation}
g(z) = \sum_{p}a_p z^p=(\sum_{k=0}^{N-1} z^k)^2= \sum_{k=0}^{N-1} (k+1) z^k +
\sum_{k=N}^{2N-2} (2N-k-1) z^k.
\end{equation}
Since $g(z)$
is analytic, then $\langle\exp[i(\phi_1 - \phi_2)\tau] \rangle=0$ if
$\tau-N\geq 0$. Otherwise, 
we can write
\begin{equation}
  \frac{1}{2 \pi i} \oint dz \, z^{\tau-N} g(z) = \frac{1}{p!} \,g^{(p)}(0),
\end{equation}
where $p = N -\tau -1$, and $g^{(p)}$ is the $p$-th derivative of $g$,
and obtain 
\begin{equation}
  \langle e^{i (\phi_1 - \phi_2) \tau}\rangle =
  \frac{-a_p}{N(N-1)} .
\end{equation}
Thus, including the contribution from the first
term at $\tau=0$, we obtain
\begin{equation}
  \langle e^{[i(\phi_j-\phi_k)\tau]}\rangle_{CUE}  = \left\{
    \begin{array}[c]{ll}
      1 &\, {\rm for}\; \tau =0 \\
      \frac{\tau-N}{N(N-1)} &\, {\rm for} \; 0 < \tau < N, \\ 
      0 &\, {\rm for}\; \tau \ge N.
    \end{array}
  \right.
  \label{eq:cue2}
\end{equation}
We now substitute the results, Eqs. (\ref{jrd14}), (\ref{jrd15}), and
(\ref{eq:cue2}) into the expression Eq. (\ref{jrd10}) the for mean
square displacement. This leads to
\begin{eqnarray}
  \langle (\Delta r)^2\rangle & = & t + t(t-1) \langle
  |J_{jj}|^2\rangle + (N-1) \langle |J_{j\neq k}|^2 \rangle
  \sum_{\tau=1}^{t-1} (t-\tau)\langle e^{i \alpha \tau} + e^{-i \alpha
    \tau}\rangle \nonumber \\
  & = & t + \frac{1}{N+1} t(t-1)  + \frac{2}{(N+1)(N-1)}  
  \sum_{\tau=1}^{M-1} (t-\tau)(\tau-N)
\end{eqnarray}
where $M=t$ for $t<N$, and $M=N$ for $t\geq N$. Carrying out the
required sums we obtain
the final result for the MSD in the CUE ensemble:
\begin{equation}
  \langle (\Delta r)^2(t) \rangle = 
  \left\{
    \begin{array}[c]{ll}    
      \displaystyle t + \frac{t(t-1)}{N+1} \left[\frac{t-2}{3(N-1)} \right]
      & \displaystyle\quad{\rm for}\; t\leq N, \nonumber \\[1em]    
      \displaystyle \frac{1}{N+1} t^2 +  \frac{N}{3} & 
      \displaystyle \quad {\rm for}\; t > N. 
    \end{array}
  \right.
\end{equation}

\subsection{The MSD in the COE ensemble}
 
\begin{figure}[htbp]
  \centering
  \includegraphics[scale=.75]{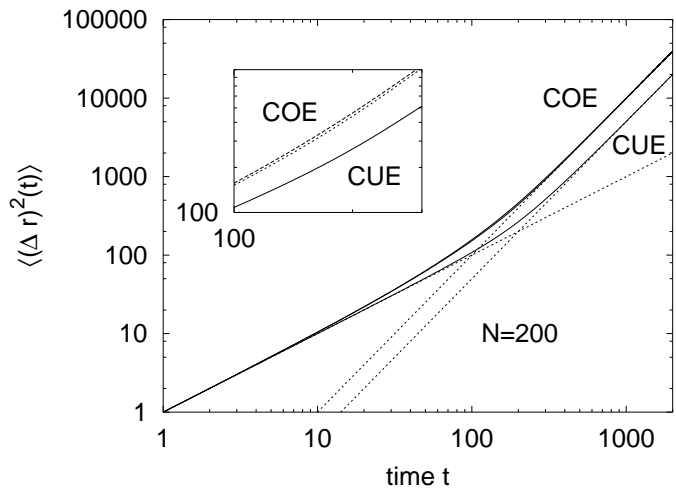}
  \caption{Log-log plot of the ensemble averages of the mean square
    displacement using RMT. Both CUE and COE results are plotted. The
    COE results are the two close curves, where the higher is the
    result given in Eq.~(\ref{eq:final}) for $k=2$, while in the lower
    curve the corrections $\delta_{<,>}$ have been neglected.  Three
    asymptotic estimates $t, t^2/N, 2t^2/N$ are also plotted.  Inset
    shows the region $t=100$ to $t=300$ where the differences between
    the two COE results are most pronounced. }
  \label{fig:rmt}
\end{figure}

The calculation of the MSD in the circular orthogonal ensemble
proceeds in very much the same way as in the CUE, the only difference
being in the pair correlation function for the quasi-energies,
$\phi_j$.  For the COE the two-point correlation function is [see
Mehta\cite{Mehta91da}, Eq.(6.2.4), Eq.(10.3.41)]
\begin{eqnarray*}
  R_2(\theta,\phi) & = & \det 
  \left[ 
    \begin{array}[c]{cc}
      \sigma_N(0) & \sigma_N(\theta-\phi) \\
      \sigma_N(\phi-\theta) & \sigma_N(0)
    \end{array}
  \right] \\ & = & 
  [(\sigma_N(0))^2]^{(0)} - [(\sigma_N(\theta-\phi))^2]^{(0)}.
\end{eqnarray*}
Here, $\sigma_N$ is a quaternion, and $\det$ is the quaternion
determinant, with 
\begin{equation}
  [(\sigma_N(\theta-\phi))^2]^{(0)} = (S_N(\theta-\phi))^2 +
  DS_N(\theta-\phi) \, JS_N(\theta-\phi). 
\end{equation}
These quantities are given by Mehta as
\begin{eqnarray}
  S_N(\theta) & = & \frac{1}{2\pi} \frac{\sin
  \frac{N\theta}{2}}{\sin \frac{\theta}{2}} =
  \frac{1}{2\pi} \sum_{k=0}^{N-1} e^{i p_k \theta} \nonumber \\
   DS_N(\theta) & = &  \frac{1}{2\pi} \sum_{k=0}^{N-1} ip_k \, e^{i p_k \theta} =
  \frac{d}{d\theta} S_N(\theta) \nonumber \\
  JS_N(\theta) & = & -\frac{1}{2\pi i} \sum_q \frac1q e^{iq\theta},
 \end{eqnarray}
where $p_k = \frac12-\frac{N}{2} + k, \quad k=0,1,\dots,N-1 $, and $q
= \pm \frac12(N+1), \pm \frac12(N+3), \dots $.  Then
\begin{equation}
  \langle e^{i (\theta-\phi) \tau} \rangle_{COE}  = \frac{1}{N (N-1)}[
  S_1(\tau) +  S_2(\tau) + S_3(\tau)],
\end{equation}
where
\begin{eqnarray}
  S_1(\tau) & = &  \int  d\theta \int d\phi \,e^{i(\theta-\phi)\tau}
  \frac{N^2}{4\pi^2}, \nonumber \\
  S_2(\tau) & = & -\int  d\theta \int d\phi \, e^{i(\theta-\phi)\tau}
  \frac{1}{4\pi^2} \sum_{k,l=0}^{N-1} e^{i(p_k + p_l)(\theta-\phi)}, \nonumber \\
  S_3(\tau) & = &  \int  d\theta \int d\phi \,e^{i(\theta-\phi)\tau}
  \frac{1}{4\pi^2} \sum_{p,q} \frac{p}{q} e^{i(p + q)(\theta-\phi)}. 
\end{eqnarray}
The evaluation of these sums and integrals is relatively
straightforward, and leads to
\begin{equation}
  \langle e^{[i \alpha \tau]}\rangle_{COE}  = \left\{
    \begin{array}[c]{ll}
      1 &\, {\rm for}\; \tau =0 \\
      \frac{1}{N (N-1)}[-N + 2\tau[f(\frac{N}{2}+\tau)-f(\frac{N}{2})] ] &\,
      {\rm for} \; 0 < \tau < N, \\  
      \frac{1}{N(N-1)}[-N + 2\tau[f(\frac{N}{2}+\tau)-f(\tau-\frac{N}{2})] ]
      &\, {\rm for}\; \tau \ge N. 
    \end{array}
  \right.
  \label{eq:coe2}
\end{equation}
Here $f(T)$ is defined by
\begin{equation}
  f(T) := \sum_{k=1}^T \frac{1}{2k-1} = 1 + \frac13 + \dots + \frac{1}{2T-1}.
\end{equation}
This function has at most a logarithmic dependence on its upper limit
for large $T$.  For the MSD in the COE we then obtain
\begin{equation}
  \langle (\Delta r)^2(t) \rangle = 
  \left\{
    \begin{array}[c]{ll}    
      \displaystyle t + \frac{t(t-1)}{N+2} \left[ 1 +
        \frac{t-2}{3(N-1)} \right] +\delta_{<} 
      & \displaystyle\quad{\rm for}\; t\leq N, \nonumber \\[1em]    
      \displaystyle  \frac{2}{N+2} t^2 +
      \frac{N}{3} - \frac{N}{3(N+2)}+\delta_{>} & 
      \displaystyle \quad {\rm for}\; t > N. 
    \end{array}
  \right.
\end{equation}
Here $\delta_{<,>}$ are small corrections to the explicit formulae
that have to be evaluated numerically.  Note that the explicit results
given here for both ensembles can be combined into the single
expression
\begin{equation}
  \langle (\Delta r)^2(t) \rangle = 
  \left\{
    \begin{array}[c]{ll}    
      \displaystyle t + \frac{t(t-1)}{N+k} \left[ k-1 +
      \frac{t-2}{3(N-1)} \right] +(k-1)\delta_{<}
      & \displaystyle\quad{\rm for}\; t\leq N,  \\[1em]    
      \displaystyle \frac{k}{N+k} t^2 +
  \frac{N}{3} - \frac{N(k-1)}{3(N+k)} + (k-1)\delta_{>} & 
      \displaystyle \quad {\rm for}\; t > N. 
    \end{array}
  \right. \label{eq:final}
\end{equation}
Those results are shown in Figure~\ref{fig:rmt} for $N=200$. The COE
results are two close curves, where the higher is the result given in
Eq.~(\ref{eq:final}) for $k=2$, while in the lower curve the
corrections $\delta_{<,>}$ have been neglected.  Three asymptotic
estimates $t, t^2/N, 2t^2/N$ are also plotted.  Inset shows the region
$t=100$ to $t=300$ where the differences between the two COE results
are most pronounced.
To our delight, RMT average lends classical diffusion as the
short-time prediction in both cases, the CUE average being ``more
classical''. It is worth emphasizing that the classical behavior
persists up to the Heisenberg times $\sim h^{-1}=N$ rather than the
Ehrenfest time $\sim \ln h^{-1} = \ln N$.  On the other hand, for
times longer than the Heisenberg time we observe ballistic motion. The
ballistic coefficient is proportional to the effective Planck constant
therefore it disappears in semi-classical limit.

In other words, fixing the time and performing semi-classical limit
($h\rightarrow 0 \equiv N \rightarrow \infty$) we obtain $\langle
(\Delta r)^2(t)\rangle = t$, which is the classical result relevant
for both the classical multi-baker and the 1D random walk, which is
modeled by the classical system.  Fixing the Planck constant
($N=const$) and performing long time limit we observe $\langle (\Delta
r)^2(t)\rangle = k t^2 / N $, which is reflection of the crystal-like
structure of the system.

\subsection{Extremal properties of the MSD}
\label{sec:extr}

To get another perspective on the above results consider a more
general class of quantum multiplexer maps~\cite{wojcik02dc} where the
local dynamics is given by a different map $B$, e.g a cat map,
standard map, or other, replacing the baker map. Then the
formula~(\ref{jrd11}) is still valid, however the eigenvalues and
eigenvectors are those of the new map $B$.

Using formula~(\ref{jrd11}) it is  easy to see that the time-dependent
mean  square  displacement for  any  quantum  multiplexer  map has  to
satisfy
\begin{enumerate}
\item $\langle (\Delta r)^2(0) \rangle = 0$
\item $\langle (\Delta r)^2(1) \rangle = 1$
\item $0 \leq \langle (\Delta r)^2(t) \rangle \leq t^2$
\end{enumerate}
In fact, those results are true in both the quantum and the classical
case. Moreover, it is possible to find local dynamics which realize
both of the extremal cases. 

To see it, take as the local map a right-left exchange operator,
defined classically by
\[
  B(n,x,y) := 
  \left\{
    \begin{array}[c]{ll}
      (n,x+1/2, y), &\quad {\rm for} \; 0 \leq x < 1/2, \\
      (n,x-1/2, y), &\quad {\rm for}\; 1/2 \leq x < 1.
    \end{array}
  \right.
\]
and quantum mechanically by
\[
\bb := \left[
  \begin{array}[c]{cc}
    0 & 1 \\ 1 & 0
  \end{array}
\right]
\]
(using the context of Section~\ref{sec:qmb}). The dynamics is obvious:
the particle jumps between two neighboring sites, leading to the mean
square displacement having values 0 for even and 1 for odd times.

On the other hand, taking identity as the local dynamics, we induce
ballistic motion: particle starting in the state going initially to
the right will keep on going to the right, leading to the ballistic
transport: 
\[
\langle (\Delta r)^2(t) \rangle = t^2
\]

Therefore, we see that translational invariance of the coupling
operator $T$ allows, in principle, for large asymptotic freedom:
\[
\langle (\Delta r)^2(t) \rangle \propto t^\alpha,
\]
where $0\leq\alpha\leq 2$~\footnote{A more precise statement is: there
  exists a constant $\alpha  \in [0,2]$ such that $\lim_{t \rightarrow
    \infty} \langle (\Delta r)^2(t)  \rangle / t^{\beta} = \infty$ for
  $\beta<  \alpha$, and $\lim_{t  \rightarrow \infty}  \langle (\Delta
  r)^2(t)  \rangle  /  t^{\beta}  =  0$ for  $  \alpha<  \beta$}.   An
interesting question is,  what can be realized in  practice.  While no
constraints seem to  be imposed on the classical  level, the structure
of  the quantum mean  square displacement,  Eq.~(\ref{eq:26}) suggests
that for fixed  $\hbar$ only $\alpha=0$ or $\alpha=2$  are viable. One
of the  questions it raises is  what conditions need to  be imposed on
the  internal  dynamics so  that  the  semi-classical  limit leads  to
anomalous diffusion $\alpha \neq 1  $.  We interpret our RMT result as
follows: internal fully chaotic dynamics (mixing) of the classical map
implies  generically diffusion.   We  are thus  led  to believe,  that
anomalous diffusion can arise  in semi-classical limit in systems with
partially  chaotic, partially  integrable internal  dynamics.  Similar
observations were made often before  in the context of different types
of systems and transport in phase space as opposed to the transport in
real  space,  which we  discuss  here.  More  precise results  require
further study.

\section{Comparison of numerical results with random matrix theory
  predictions}  
\label{sec:numerics}

We have studied numerically a number of cases of quantum multi-bakers
as well as some more general systems~\cite{wojcik02dc}. Here we
present some of those results to show the quality of prediction
afforded by the random matrix theory.

Our numerical evaluation of the formula~(\ref{jrd11}) shows that
generically, almost every choice of phases defining the
quantization~(\ref{eq:3}) leads to results between the COE and CUE
results. 
\begin{figure}[htbp]
  \centering
  \includegraphics[scale=.5]{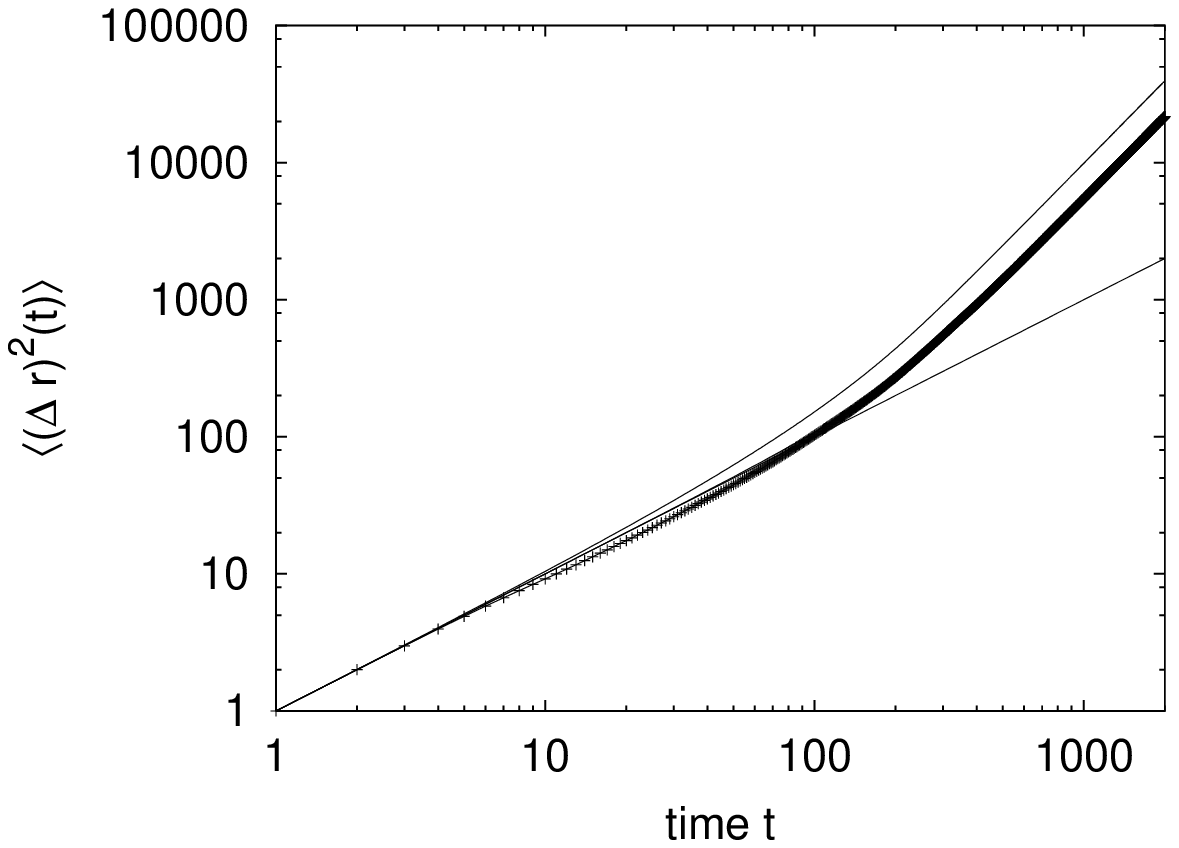} 
  \includegraphics[scale=.5]{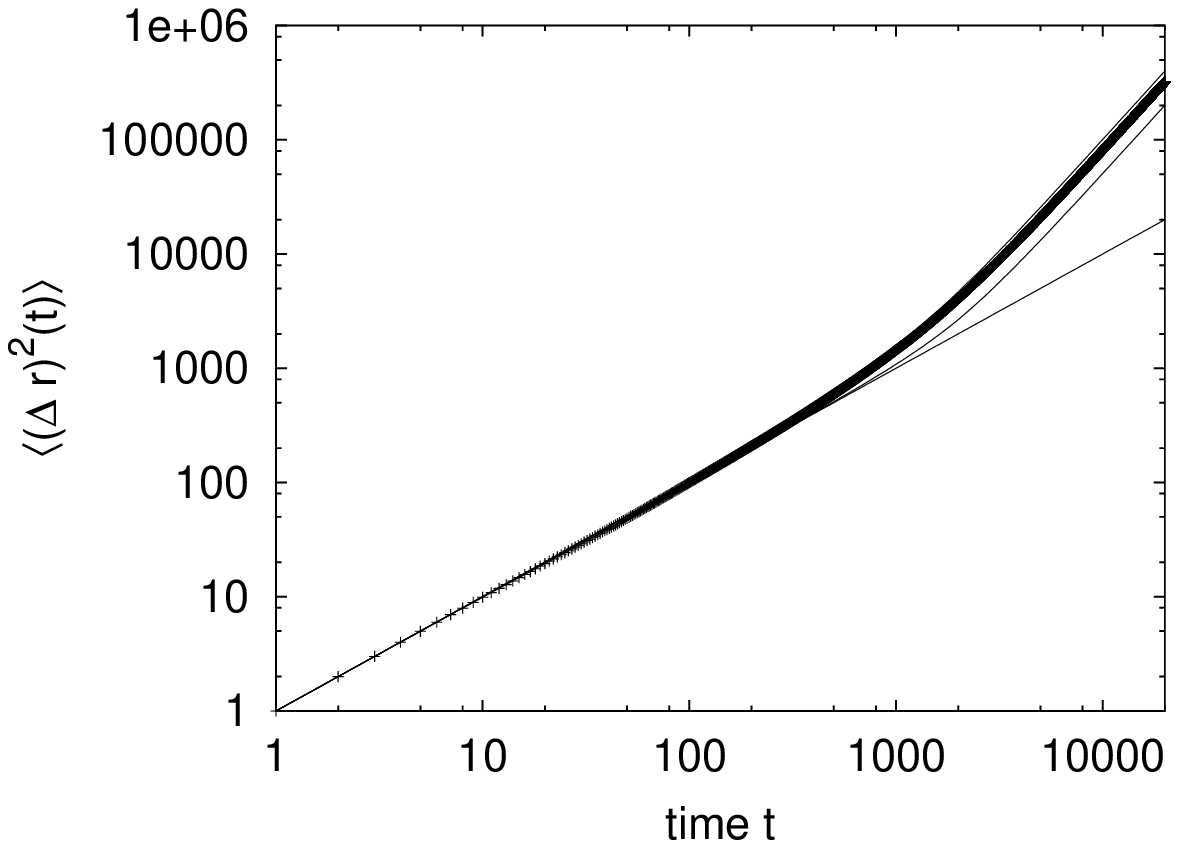}
  \includegraphics[scale=.5]{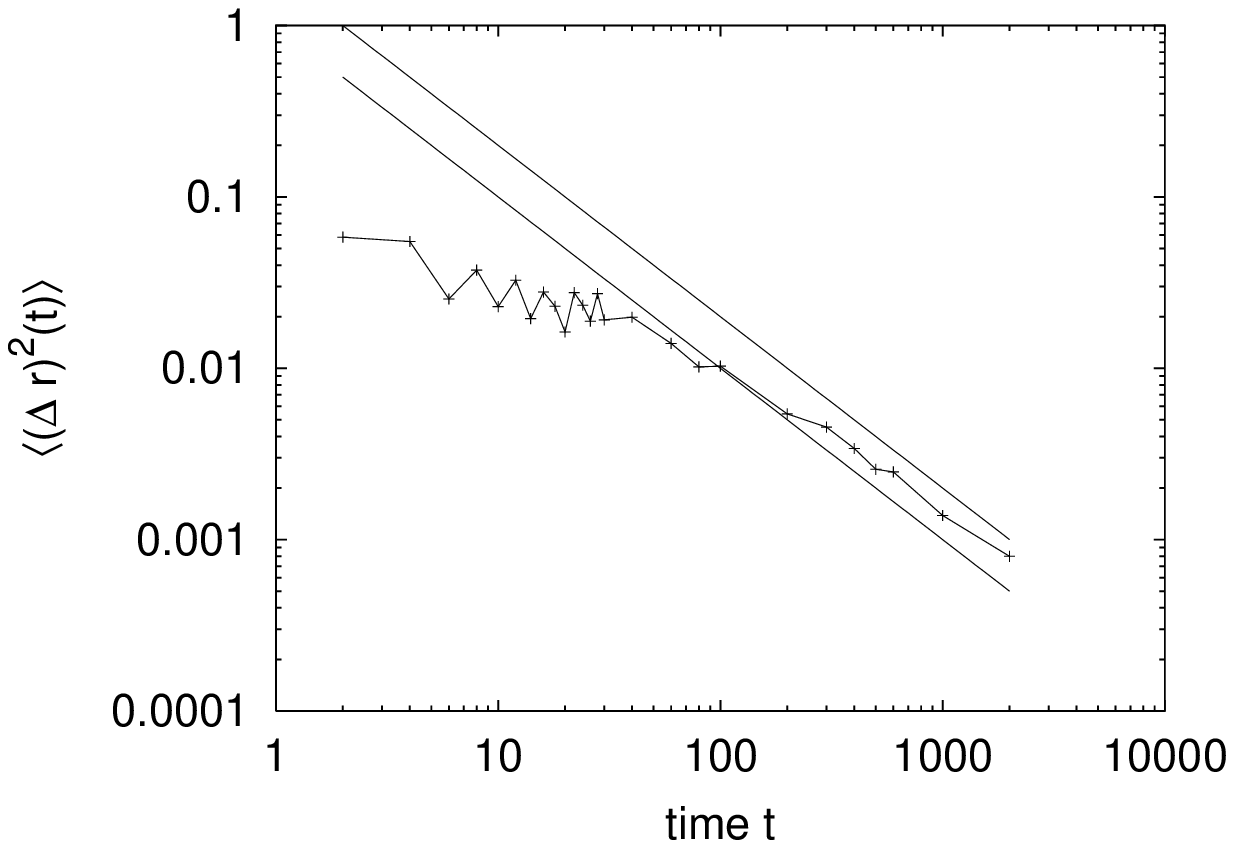}
  \caption{Log-log plot of the mean square displacement in the quantum
    multi-baker~(\ref{eq:qmb}) with phases $\phi_q=0.73,\,
    \phi_p=0.11$ as a function of time for $N=200$ (top left) and for
    $N=2000$ (top right). The crossover between the diffusive and
    ballistic behavior is clearly visible on both plots at the
    Heisenberg time $t=N$. The bottom figure shows the values of
    $\frac1N \sum_j \langle |J_{jj}|^2 \rangle$ calculated for the
    same quantum multi-baker for various values of $N$ with the
    predictions of random matrix theory~(\ref{jrd14}). We see that for
    $N > 50 $ the results lie between the analytic values for COE and
    CUE ensembles.}
  \label{fig:qmb1}
\end{figure}
Figure~\ref{fig:qmb1} shows results obtained for the quantum
multi-baker~(\ref{eq:qmb}) with phases $\phi_q=0.73,\, \phi_p=0.11$.
We show the mean square displacement as a function of time for $N=200$
and for $N=2000$ on log-log plot. There is a clear crossover between
the diffusive and ballistic behavior on both plots.
Figure~(\ref{fig:qmb1}.c) compares the values of $\frac1N \sum_j
\langle |J_{jj}|^2 \rangle$ calculated for the same quantum
multi-baker for various values of $N$ with the predictions of random
matrix theory~(\ref{jrd14}). We see that for $N > 50 $ the results lie
between the analytic values for COE and CUE ensembles.

\begin{figure}[htbp]
  \centering
  \includegraphics[scale=.5]{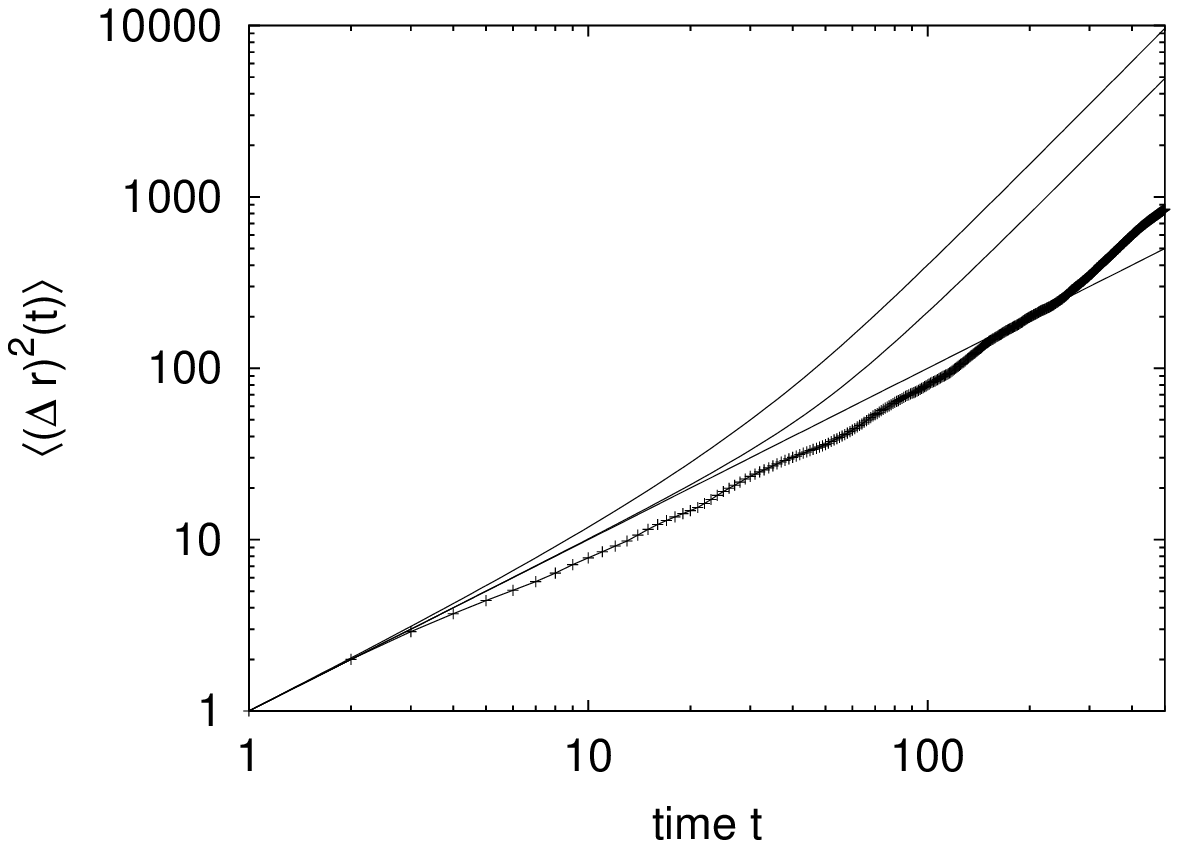}  
  \includegraphics[scale=.5]{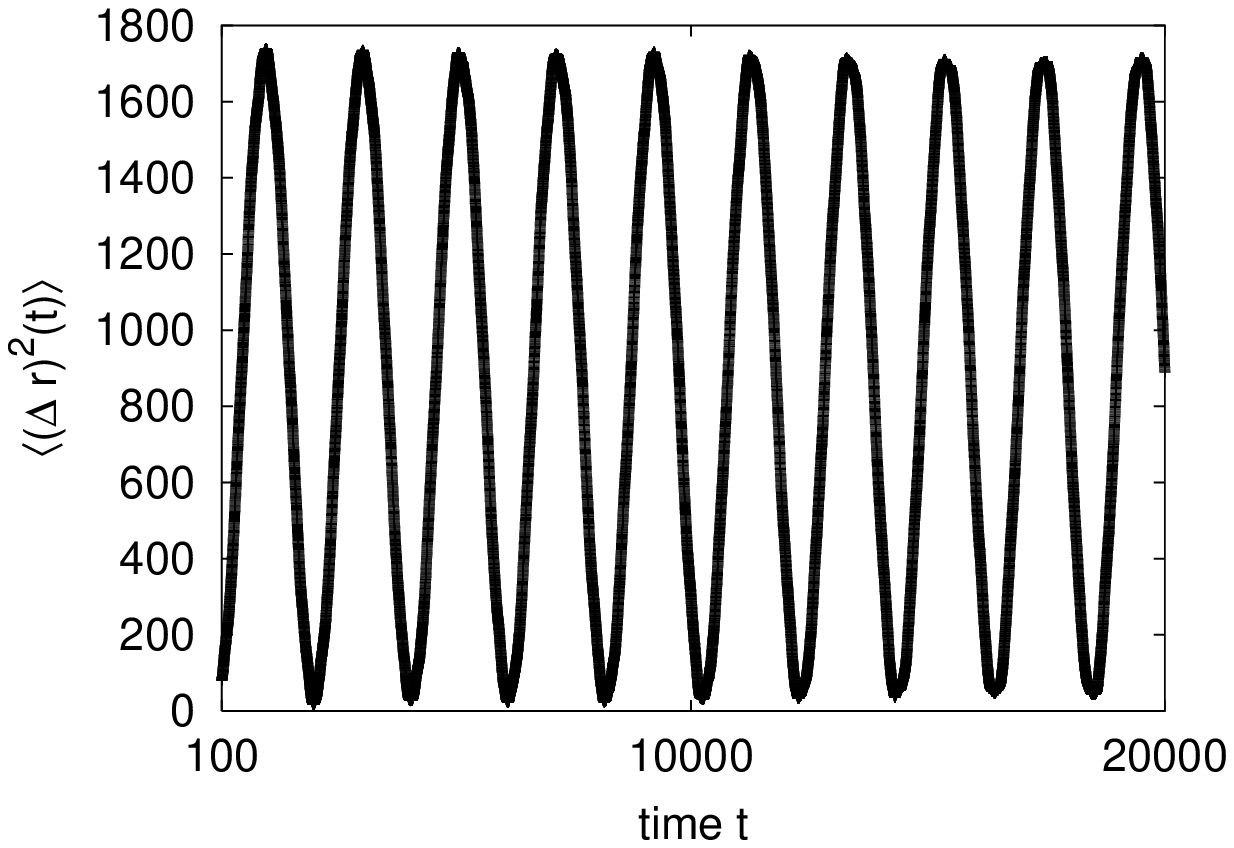} \\ 
  \includegraphics[scale=.5]{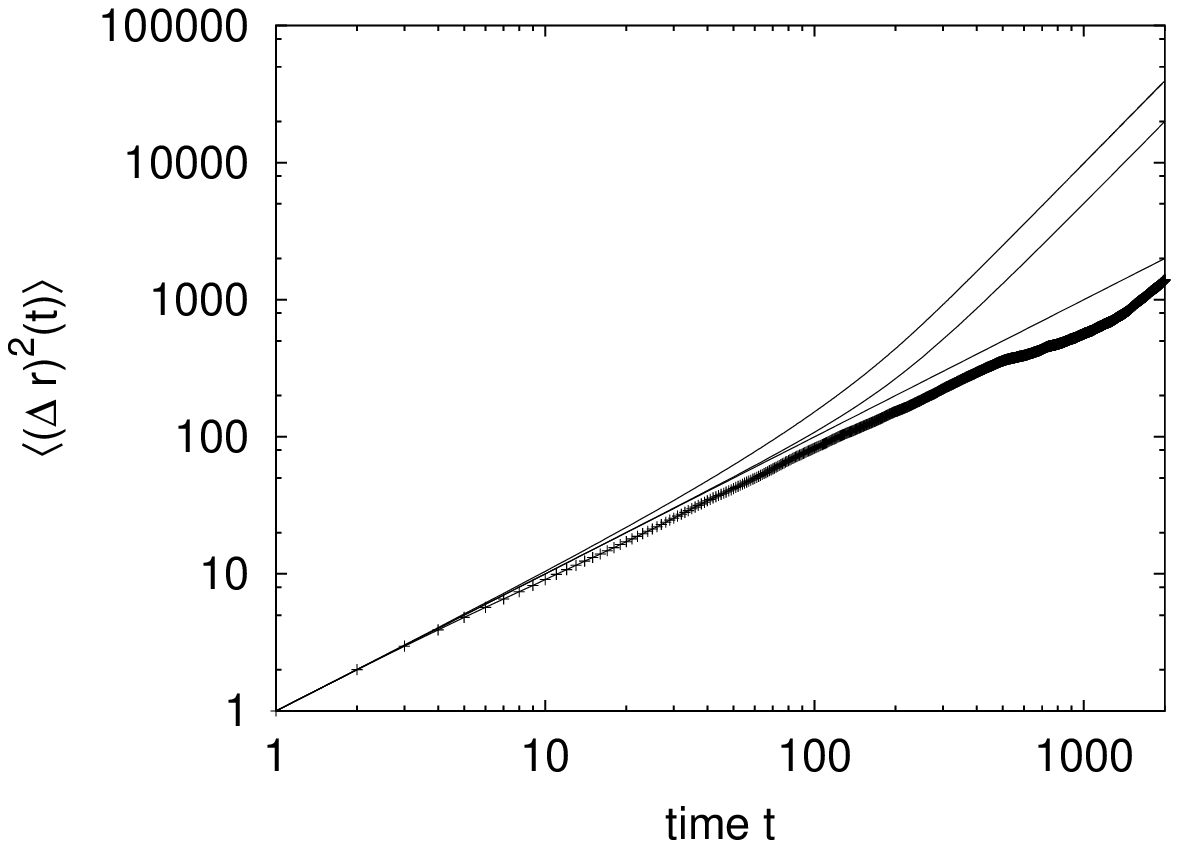}  
  \includegraphics[scale=.5]{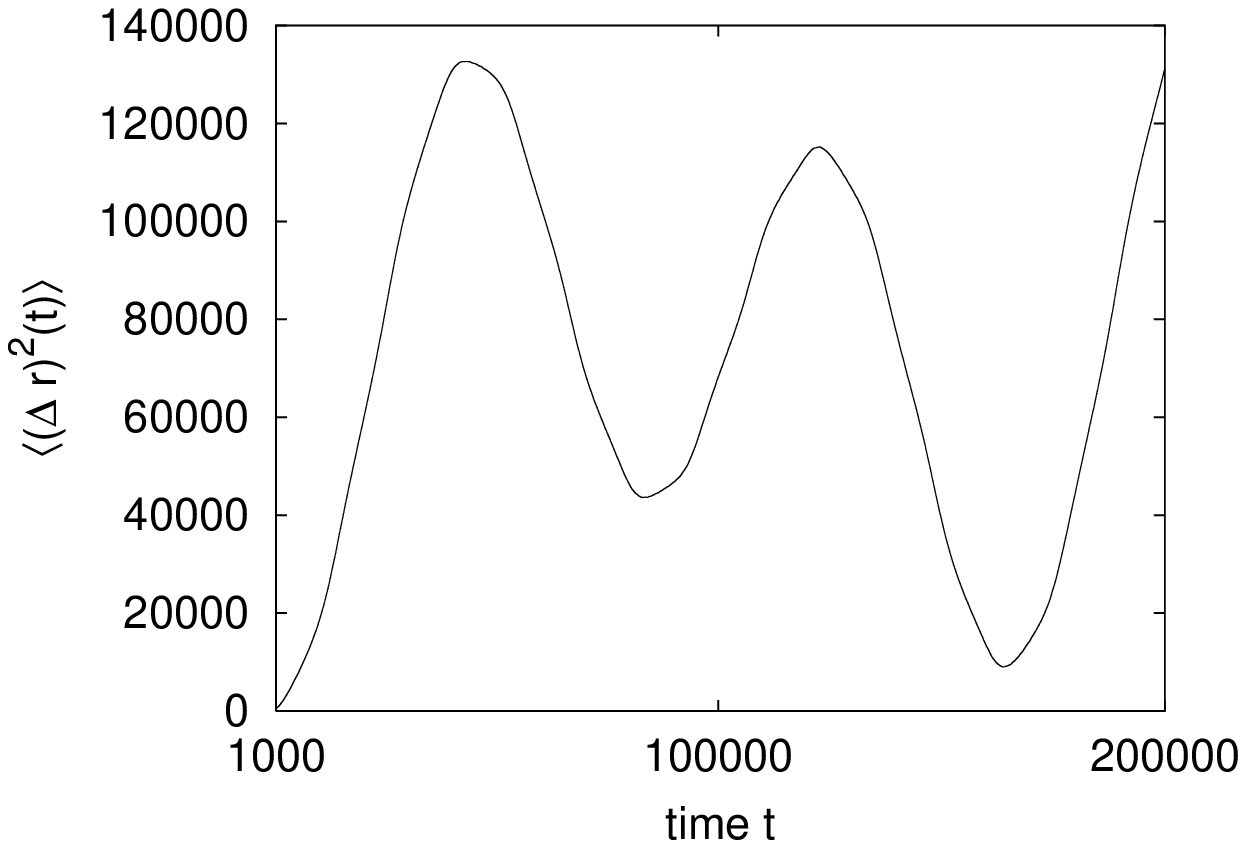}
  \caption{Mean square displacement for quantum multi-baker with phases
    $\phi_q=\phi_p=1/2$ for $N=50$ (top row) and $N=200$ (bottom row).
    The left plots show the short time behavior on the log-log scale.
    We see the same diffusive behavior as in the generic case. The
    right plots are normal scale and show the oscillations encountered
    on the long scale. All the plots are numerical evaluations of the
    exact formula~(\ref{eq:26}).}
  \label{fig:qmb2}
\end{figure}

In cases when the two phases add up to one $\phi_q + \phi_p =1$, we
find from our numerical work that the ballistic coefficient $\frac1N
\sum_j \langle |J_{jj}|^2 \rangle$ vanishes. Therefore, only the
oscillating contribution remains in the expansion~(\ref{eq:26}). We
still observe diffusive behavior up to Heisenberg time, although not
as clearly as for generic systems. After this transient the particle
localizes and the mean square displacement oscillates irregularly
around some average value. Figure~\ref{fig:qmb2} shows the mean square
displacement for quantum multi-baker with phases
$\phi_q=\phi_p=1/2$~\cite{Saraceno90} for $N=50$ (top row) and $N=200$
(bottom row). In both cases the left plot shows the short time
behavior on the log-log scale. We see the same diffusive behavior as
in the generic case. The right plots are normal scale and show the
oscillations encountered on the long scale. All the plots are
numerical evaluations of the exact formula~(\ref{eq:26}).  We suspect
that this non-generic, localized behavior will be seen whenever the
phases $\phi_q,\phi_p$ sum to unity, but both the proof and a physical
understanding remain to be developed.

\section{Multi-baker maps and quantum random walks}
\label{sec:qrw}

We conclude this discussion of the quantum multi-baker maps by
establishing connections between our work and some other research on
quantum random walks. We restrict ourselves to discrete quantum walks
on line.  Continuous quantum random walks~\cite{childs01f,childs02f,tamon02f}
and discrete quantum walks on graphs~\cite{aharonov01f,leroux02f} are less
connected to our work.

A popular model of a quantum random walk, known today as the {\em
  Hadamard walk}, was described some ten years ago by S.~Godoy and
S.~Fujita~\cite{GodoyF92}. It was obtained through approximating the
evolution of some special wave packets in Kr\"onig-Penney type
potential.  A similar model was discussed by D.~Aharonov, et
al.~\cite{aharonov93s}, who considered the motion of spin-$\frac12$
particles in one dimension, and proposed experimental realization of
the walk in the framework of quantum optics.
Further, D. Meyer studied quantum cellular automata for their possible
  applications to
quantum computing. In section 5 of~\cite{Meyer96}, Meyer obtained
a quantum random walk model closely related to the Hadamard walk,
  discussed later by A.~Ambainis et
al.~\cite{ambainis01f}.

Looking for an analytically tractable model of quantum transport, we
quantized~\cite{wojcik02da} the classical multi-baker map, a
deterministic model of classical random walk, obtaining a family of
quantum random walk models, the simplest case being the Hadamard walk.
We studied both the translationally invariant and random versions of
the system.

Since then a number of other papers appeared on the
topics, e.g.~\cite{bach02f,duer02f,kendon02f,kendon02fa,konno02fa,konno02fb,%
konno02fc,konno02fd,MackayBSS02,shenvi02f,travaglione02f,yamasaki02f}.
Of particular interest for our work is the work of Brun et
al.~\cite{brun02fa,brun02fb,brun02fc} who also studied the transition
to classical behavior, but instead of considering semi-classical limit
they assume interaction with environment implemented in decoherent
coins.

Much of the work on discrete quantum random walks
can easily be reformulated as a special case of the multi-baker maps
(or multiplexer maps, see ref.~\cite{wojcik02dc}) in particular, the
map with the largest possible value of Planck's constant, $h=1/2$, or
equivalently, $N=2$. In order to make this connection we review, very
briefly, the structure of quantum random walks as described by much of
the literature.

The simplest quantum random walk described in the literature is the
{\it Hadamard walk}. Here one considers a collection of quantum
states, which we will denote as $|n,d\rangle$. The parameter $n$ will
take on all integer values within a prescribed range, and here we take
the range to be ${\bf{Z}}$, all integers, positive and negative. The
parameter $d=r,l$ denotes a direction, to the {\it right} or to the
{\it left}, respectively. Next, one defines an elementary quantum
transformation, $\bh$, which takes $|n,d\rangle$ into symmetric and
anti-symmetric combinations, as
\begin{equation}
\left( \begin{array}{c}
   |n,r'\rangle \\
   |n,l'\rangle \end{array}\right) = \bh \left( \begin{array}{c}
   |n,r\rangle \\
   |n,l\rangle \end{array}\right) = \frac{1}{\surd
   2}\left( \begin{array}{cc}
   1 & 1 \\
   1 & -1 \end{array} \right) \cdot \left(\begin{array}{c}
   |n,r\rangle \\
   |n,l\rangle \end{array} \right).
\label{hw}
\end{equation}
This step is now coupled to a translation, ${\bf{T}}$ of
$|n,r'\rangle$ one unit to the right and, similarly, $|n,l'\rangle$
one unit to the left, as
\begin{eqnarray}
\bt |n,r\rangle & = & |n+1,r\rangle \nonumber \\
\bt |n,l\rangle & = & |n-1,l\rangle
\label{hwt}
\end{eqnarray}
$H$ is called the {\em Hadamard gate\/} in quantum computing
literature~\cite{Preskill99f,Nielsen00f}, which gives rise to the name
{\em Hadamard walk\/}. Finally the two operations are combined to form
the operator $ \bt\circ \bh =\bw$, which describes one step of the
Hadamard walk,
\begin{eqnarray}
 |n,r,t\rangle & = & \frac{1}{\surd 2}\left[|n-1,r,t-1\rangle +
 |n+1,l,t-1\rangle \right] \nonumber \\
 |n,l,t\rangle & = & \frac{1}{\surd 2}\left[|n-1,r,t-1\rangle -
 |n+1,l,t-1\rangle \right].
\label{hwf}
\end{eqnarray}
This is identical to Eq. (\ref{eq:qmb}) for $N=2$ for the special case
that the Balazs-Voros phases~\cite{BalazsV89}, $\phi_q =\phi_p =0$,
are used.  Therefore, the multi-baker map and the Hadamard walk are
identical, for the Balazs-Voros phases and $N=2$. The Hadamard
equations have been solved in detail for various boundary
conditions~\cite{wojcik02da,ambainis01f,bach02f,konno02fa,yamasaki02f}.
Similarly, the $N=2$ multi-baker map has been solved for a variety of
phases and boundary conditions~\cite{wojcik02da}.  It should be
appreciated that the multi-baker maps represent generalizations of
quantum random walks in a number of respects. The inclusion of phases
in the transformation equations allows one to treat a variety of maps
including uniform, periodic, quasi-periodic, and random systems,
something typically not considered in the theory of quantum random
walks. Thus one can find a range of phenomena in quantum multi-baker
ranging from localization to ballistic motion. Such phenomena show up
in many condensed matter systems. Further, the use of a variable
Planck constant allows for several channels of motion to be taking
place at once, and allows us to treat semi-classical as well as strong
quantum versions of these maps. This is what makes the multi-baker
maps so appealing for studying exact transport properties of condensed
quantum systems.

One common feature of different quantum random walks and quantum
multi-baker maps is the ability to include various models of
decoherence processes. In the multi-baker maps one should distinguish
between quenched disorder and annealed disorder. In quenched disorder
the phases $\phi_q,\phi_p$ vary randomly for cell to cell, but once
values are given for each cell, the phases keep those values for all
time steps. This allows some memory to be generated by the dynamics
and results in localization~\cite{wojcik02da}. For annealed disorder,
the phases are chosen randomly for each cell and at each time step.
The phases also may be specified up to some additive noise
contribution.  These processes represent the effects of random
external noise and result in some degree of
decoherence~\cite{kendon02f,brun02fa}. Further, in both quantum random
walk models and multi-baker maps one can produce decoherence by
replacing the unitary matrix describing the time evolution by a {\it
  completely positive} super-operator.  This allows for both
dissipation and decoherence. As this topic requires a separate
discussion, we will not pursue it here, but leave the discussion of
decoherence for further publications.

\section{Discussion of the results and other comments}

This paper described calculations of important properties of the
quantum multi-baker map, a simple model for transport in quantum
systems. This map is the quantum version of a well known, classically
chaotic system. In particular we were able to show, using random
matrix theory, that it is possible to obtain an analytic expression
for the equilibrium mean square displacement of a particle whose
motion is governed by this quantum map. This expression is of
particular importance for an understanding of transport in quantum
systems because it allows us to carefully examine the particle's
average motion both for finite times and for non-zero values of
Planck's constant. Thus we can see how the two non-commuting limits,
$t\rightarrow \infty$ or $\hbar \rightarrow 0$, interact with each
other. While our results are not surprising, in the sense that we
already knew qualitatively the limiting forms, we did not have a
general expression for the mean square displacement even for simple
quantum systems before this work was carried out. Moreover, the
results from random matrix theory agreed very well with careful
numerical evaluations of the mean square displacement for the
multi-baker map. This agreement is also very important because random
matrix theory does not really apply to any individual system, as far
as one knows, but rather to the average behavior of an ensemble of
similar systems. Thus in the absence of a rigorous justification of
the use of random matrix methods, their predictions must always be
checked by comparisons with numerical studies.

It will be interesting to find other signatures of classical behavior
in the quantum multi-baker map that are more closely connected to the
chaotic behavior of the classical map than the mean square
displacement. For example, is there any trace of the classical,
hyperbolic dynamical behavior in the quantum version of the map? One
prominent feature of the classical multi-baker model is the formation
of fractal structures on arbitrarily fine scales as an initial
non-equilibrium distribution of points relaxes to a final uniform
equilibrium. This fractal structure can, in turn, be related to a
positive entropy production typical of the approach to equilibrium of
a macroscopic system. The quantum version of this phenomenon is
largely unexplored and is of considerable interest to us.

Further work will also be devoted to exploring the consequences of the
freedom one has in choosing phases of the unitary operators. This
freedom allows us to consider random, quasi-random, and periodic
systems, each with its own interesting physical properties. The
quenched random system has been explored to some
extent~\cite{wojcik02da}, and provides a simple model of localization
in one-dimensional disordered systems.  Here the classical limit is
expected to be recovered through a phenomenon where the localization
length grows infinitely large as Planck's constant approaches zero,
such that almost normal diffusion takes place within each localized
region. This picture needs to be verified, and further systems still
need to be explored.

We also began to establish connections between this work on quantum
multi-baker maps and a body of related work on quantum random walk
processes. Much of that literature is, in fact, devoted to the
simplest multi-baker model, although this connection has not been
recognized in the random walk literature. Here we showed that the
$N=2$ quantum multi-baker process is identical to the Hadamard walk of
random walk theory. Although quantum computing is still far from being
realized~\cite{Preskill99f,Nielsen00f}, it is possible that the theory
of quantum random walks may be of some value for constructing
algorithms based on quantum walks that will be faster than current
classical random walk algorithms~\cite{childs02f,shenvi02f}.  Future
work will be devoted to expanding the connections between the quantum
random walks and multi-baker processes for a variety of model systems.

\section{Acknowledgments}

The  authors  would  like  to   thank  R.   Blume-Kohout  and  Profs.  
S.~Fishman, P.~Gaspard, M.~Ku\'s,  S.~Tasaki, J.~Vollmer, W.  \.Zurek,
and  K.~\.Zyczkowski for  helpful comments.   DKW is  grateful  to the
Institute for Physical Science and Technology, University of Maryland,
where  most  of  this work  has  been  done,  for support  during  his
postdoctoral stay; and to the  Center for Nonlinear Science, School of
Physics, Georgia Institute of Technology  for support as a Joseph Ford
Fellow. He  also thanks the organizing committee  of the International
Workshop   and  Seminar   on  Microscopic   Chaos  and   Transport  in
Many-Particle Systems, Dresden, 2002 for giving him the opportunity to
present this work.  JRD acknowledges support from the national Science
Foundation under Grant PHY-98-20824.

\end{document}